\documentclass[trackchanges, twocolumn]{aastex701}
\usepackage{amsmath, amssymb, bm, multirow}

\makeatletter
\AtBeginDocument{%
 \appdef\tabular@hook{\spaceskip=1sp\relax}%
 \appdef\endtabular@hook{\spaceskip=\z@skip}%
 \appdef\cell@font{\spaceskip=\z@skip}%
}
\makeatother

\usepackage{placeins}

\colorlet{GREEN}{green}

\begin{document}

\title{Beyond the Damped Random Walk: Multiband AGN Variability from SDSS to Rubin LSST}

\correspondingauthor{Hyungsuk Tak}\email{tak@psu.edu}

\author{Izak Schmidlkofer}
\affiliation{Department of Statistics, Oregon State University, Corvallis, OR 97331, USA}
\email{schmidiz@oregonstate.edu}
\author{Zhirui Hu}
\affiliation{Department of Statistics, Oregon State University, Corvallis, OR 97331, USA}
\email{Zhirui.Hu@oregonstate.edu}
\author{Weixiang Yu}
\affiliation{Department of Astronomy and Astrophysics,  Pennsylvania State University, University Park, PA 16802, USA}
\email{wuy10@psu.edu}
\author[0000-0003-0334-8742]{Hyungsuk Tak}
\affiliation{Department of Astronomy and Astrophysics,  Pennsylvania State University, University Park, PA 16802, USA}
\affiliation{Department of Statistics,  Pennsylvania State University, University Park, PA 16802, USA}
\affiliation{Institute for Computational and Data Sciences,  Pennsylvania State University, University Park, PA 16802, USA}
\affiliation{The Research Institute of Basic Sciences, Seoul National University, Seoul 08826, Republic of Korea}
\email{tak@psu.edu}
%\affiliation{Institute for Gravitation and the Cosmos, The Pennsylvania State University, University Park, PA 16802, USA}


\begin{abstract}

The flux variability of Active Galactic Nuclei (AGNs) is commonly
modeled using the damped random walk (DRW), but increasingly rich
multiband time-domain surveys motivate more flexible stochastic
models that characterize variability beyond a single timescale and
exploit dependence across bands. Multivariate continuous-time
autoregressive moving average (MCARMA) processes extend the
multivariate DRW to higher-order stochastic dynamics and provide a
matrix-valued power spectral density for characterizing marginal
variability and cross-band dependence. In a carefully selected Sloan
Digital Sky Survey (SDSS) Stripe~82 sample, higher-order models are
selected for 109 of 134 AGNs (81.3\%), with substantial rather than
marginal support in most objects; convergence diagnostics indicate
that this fraction is conservative. The fitted models also reveal
strong cross-band dependence that generally decreases with wavelength
separation. Motivated by this richer structure, we investigate its
recoverability under Vera C. Rubin Observatory Legacy Survey of Space
and Time (LSST) observing conditions. Model-order recovery depends
strongly on the underlying stochastic behavior as well as source
brightness and observing cadence. Recovery is high across many of the
simulated settings, but deteriorates when the characteristic features
of higher-order variability become difficult to distinguish from the
multivariate DRW. Additional simulations show that this loss of
distinguishability can persist over a range of underlying variability
behaviors. These results show that stochastic structure beyond the
multivariate DRW is prevalent in our selected SDSS sample, while its
recoverability with LSST depends on both observational conditions and
the form of the underlying stochastic variability. The methodology is
implemented in the Python package \texttt{mcarma}.

\end{abstract}


%\keywords{\uat{Cosmology}{343} --- \uat{High Energy astrophysics}{739} --- \uat{Interstellar medium}{847} --- \uat{Stellar astronomy}{1583} --- \uat{Solar physics}{1476}}


\section{Introduction}
\label{sec1}

The damped random walk (DRW) process \citep{kelly2009variations},
also known as the Ornstein--Uhlenbeck process or a continuous-time
autoregressive process of order one, is among the most widely used
stochastic models for active galactic nucleus (AGN) variability.
The DRW is characterized by a single damping timescale separating
correlated short-timescale variability from effectively decorrelated
long-timescale behavior, with its amplitude setting the overall
variability level. Its stochastic parameters have been related to
astrophysical properties such as black hole mass and luminosity
\citep{kelly2009variations, macleod2010modeling, kozlowski2010quantifying,
kim2012, andrae2013}, and the model has also been widely used in
time-delay and reverberation-mapping analyses
\citep{2015ApJ...799..168D, 2015ApJ...800...11L, tak2017bayesian,
zu2011an, zu2016app, 2024ApJS..272...26S}.

The limitations of the DRW have also been well documented.
\cite{2011ApJ...743L..12M} and \cite{2013ApJ...765..106Z} found
departures from DRW behavior on short timescales, while
\cite{2014MNRAS.439..703G} and \cite{2015MNRAS.451.4328K} showed
that a single characteristic timescale can be too restrictive for
describing the diversity of quasar variability, including
quasi-periodic behavior and multiple characteristic timescales
\citep[e.g.,][]{2023ApJ...950...37M}. More generally, fitting a DRW to a
process with different stochastic dynamics can distort the inferred
timescale and complicate its physical interpretation
\citep{2016MNRAS.459.2787K, 2017A&A...597A.128K,
tak2017bayesian}. As astronomical time-domain data become longer,
denser, and more precise, practical stochastic models that extend
beyond the DRW are therefore increasingly important for detailed
characterization of AGN variability.

These considerations have motivated the use of the more flexible
continuous-time autoregressive moving average (CARMA) family
\citep{kelly2014flexible}. CARMA processes generalize the DRW to
accommodate multiple characteristic timescales and a broader range of
stochastic dynamics, including damped and quasi-periodic oscillations,
while producing flexible power spectral densities (PSDs) with
multiple characteristic frequencies, breaks, and slopes. Their
frequency-domain properties have made them useful for characterizing
stochastic variability in a broad range of AGN light curves
\citep[e.g.,][]{kelly2014flexible, 2019PASP..131f3001M,
2022ApJ...936..132Y, 2022MNRAS.514..164S, 2025ApJ...992..130Y}. Existing astronomical
applications of CARMA, however, have largely analyzed individual
photometric bands separately.

Joint modeling across bands is increasingly important in time-domain
astronomy because correlated light curves contain shared information
that can improve inference when individual bands are sparsely sampled
or noisy, while also providing direct inference on cross-band
dependence. Within the CARMA framework, astronomical multivariate
modeling has largely been restricted to the multivariate DRW
\citep{2020AJ....160..265H}, which retains first-order DRW dynamics
within each band. \citet{Schmidlkofer2026} recently adapted
Wiener-driven multivariate CARMA (MCARMA) processes \citep{tomasson2018} to astronomical
multiband light curves, extending the multivariate DRW to higher-order
band-specific stochastic dynamics while retaining joint modeling of
cross-band dependence. They develop the statistical methodology and estimation procedure.
Through controlled simulations, they demonstrate that joint
multivariate modeling can substantially improve parameter and
spectral estimation, while also identifying regimes in which
higher-order dynamics are weakly identifiable.

In this work, we use the MCARMA framework to investigate richer
multiband stochastic behavior in observed AGN light curves and its
recoverability under the observing conditions of the Vera C. Rubin
Observatory's Legacy Survey of Space and Time
\citep[LSST;][]{ivezi2019lsst}. We first analyze five-band $ugriz$
light curves from SDSS Stripe~82
\citep{2000AJ....120.1579Y}, drawn from the 9,258 spectroscopically
confirmed quasars in the SDSS Data Release Seven quasar catalog
compiled by \citet{macleod2012}. After requiring well-sampled light
curves and screening for stationarity, we obtain a final sample of
134 AGNs. Their long observational baselines and nearly simultaneous
five-band measurements make this sample particularly informative for
investigating both marginal stochastic variability and cross-band
dependence.

Among the three candidate orders considered, i.e., MCARMA$(1,0)$,
MCARMA$(2,0)$, and MCARMA$(2,1)$, the SDSS analysis reveals
substantial stochastic structure beyond the multivariate DRW.
Higher-order MCARMA models are selected for 109 of the 134 AGNs
(81.3\%), including 105 objects (78.4\%) for which MCARMA$(2,1)$,
the most flexible order considered, is selected. As discussed in
Section~\ref{sec:sdss.select}, incomplete optimization can only make
the 81.3\% higher-order fraction conservative rather than inflate it.
For most objects selecting a higher-order model, the preference over
lower-order alternatives is substantial rather than marginal. The
selected second-order models are predominantly overdamped, indicating
that the additional variability structure is generally characterized
by multiple relaxation modes rather than damped oscillatory behavior.
Across the selected SDSS sample, the fitted marginal PSDs show
departures from the DRW spectral shape, with relatively greater
variability power on long timescales, while the inferred cross-band
dependence is strongest between neighboring bands and generally
decreases with increasing wavelength separation.  Together, these findings indicate richer
stochastic behavior than the multivariate DRW in the selected SDSS
sample, involving both diverse within-band dynamics and strong
dependence across bands.

The SDSS findings naturally raise an important question for LSST: can
this richer multiband stochastic behavior still be identified and
characterized under substantially different observing conditions? In
contrast to the nearly simultaneous five-band measurements in our
selected SDSS sample, LSST observations are staggered across filters
and irregularly distributed in time. LSST will also extend variability
studies to substantially fainter sources, for which reduced
photometric precision can make stochastic structure more difficult to
recover. The SDSS sample considered here is concentrated around
\(20\) mag, with median object-level magnitudes ranging from 19.57 in
the \(z\) band to 20.44 in the \(u\) band.

We therefore simulate realistic Wide-Fast-Deep (WFD) and Deep Drilling
Field (DDF) observing sequences at source magnitudes 19.5 and 22.0
while holding the intrinsic variability amplitude fixed. The WFD
survey covers approximately 18,000 deg$^2$ of the southern sky,
providing a total of about 800 visits across the six $ugrizy$
photometric bands over 10 years, whereas the DDF survey consists of
six extragalactic fields covering a total of approximately 60 deg$^2$
and provides at least an order of magnitude more visits than WFD.
The 19.5-mag setting approximately represents the brightness regime
of our SDSS sample, whereas the 22.0-mag setting probes the
substantially fainter regime that LSST will make accessible. Together, the two brightness levels and two observing strategies allow us to
examine how photometric precision and sampling density affect recovery
of different forms of stochastic dynamics within bands and dependence
across bands.

The LSST simulations show that model-order recovery depends not only
on photometric precision and sampling density, but also strongly on
the form of the underlying stochastic variability. Higher-order
dynamics are reliably distinguished from the multivariate DRW across
many of the simulated settings, whereas recovery deteriorates when
their characteristic variability features become less distinguishable
from lower-order behavior. Additional simulations show that this
difficulty persists across a range of nearby stochastic regimes rather
than being confined to a single parameter configuration. These results
emphasize that an apparently DRW-like light curve need not imply
intrinsically first-order variability: higher-order stochastic
structure can become difficult to identify under particular intrinsic
and observational conditions.

We further examine the recovered marginal PSDs and cross-band
coherence as scientifically interpretable summaries of the joint
parameter estimation. The marginal PSDs show how errors in the
inferred stochastic dynamics propagate to frequency-dependent
variability, with joint multiband fitting particularly beneficial
when individual bands are information-limited. Cross-band coherence
provides a complementary picture of the dependence between bands.
Its decline with increasing wavelength separation is recovered
across the simulated settings, with DDF sampling producing
substantially tighter recovery than WFD sampling. Unlike the
marginal PSDs, this cross-band information is unavailable from
independent single-band fits. Together with the SDSS results,
these simulations show how the underlying stochastic behavior,
observing design, photometric precision, and information shared
across bands determine our ability to reveal and characterize
richer stochastic variability beyond the multivariate DRW.

The remainder of the paper is organized as follows.
Section~\ref{sec2} briefly summarizes the MCARMA framework and the
likelihood function, estimation, and model-selection procedures used in our
analyses. Section~\ref{sec:sdss} describes the SDSS Stripe~82 sample
and presents the resulting stochastic and cross-band variability
properties. Section~\ref{sec:lsst} describes the LSST simulations and
investigates recovery of higher-order multiband variability under WFD
and DDF observing conditions. Section~\ref{sec:conclusion} discusses
the astronomical implications of our findings, limitations of the
current analysis, and directions for future studies of multiband
variability.

\section{MCARMA Model and Inferential Framework}
\label{sec2}

The statistical formulation and properties of the MCARMA framework
used in this work are developed in detail by
\citet{Schmidlkofer2026}. Here we provide only the elements needed
for the astronomical analyses and interpretation below. In
particular, we briefly describe the structured MCARMA model and its
frequency-domain representation, the state-space likelihood for
irregular and partially observed multiband light curves, and the
estimation and model-selection procedure used throughout this work.

Our emphasis is therefore on how the framework accommodates the
observational characteristics of the SDSS and LSST data and how its
fitted quantities can be interpreted astronomically, rather than on
the underlying statistical derivations. Detailed results concerning
the state-space formulation, parameterization and identifiability,
numerical optimization, uncertainty quantification, and controlled
simulation performance are given in \citet{Schmidlkofer2026}.

\subsection{MCARMA Model and Spectral Representation}
\label{sec2.1}

The $k$-dimensional zero-mean MCARMA$(p,q)$ process
$\mathbf{X}(t)=(X_1(t),\ldots,X_k(t))^\top$, with autoregressive
(AR) order $p$ and moving-average (MA) order $q$, is defined by
\begin{equation}
\label{eq:sde.mcarma}
A(D)\mathbf{X}(t)=M(D)\,d\mathbf{B}(t),
\end{equation}
where $p>q\geq0$, $D=d/dt$, and
\begin{equation*}
A(D)
=
I_kD^p+A_{p-1}D^{p-1}+\cdots+A_0
\end{equation*}
and
\begin{equation*}
M(D)
=
M_qD^q+\cdots+M_1D+I_k
\end{equation*}
are the AR and MA matrix polynomials, respectively. Here $I_k$
denotes the $k\times k$ identity matrix, and
$A_i,M_i\in\mathbb{R}^{k\times k}$ are coefficient matrices. The
leading AR coefficient and constant MA coefficient are normalized to
$I_k$. A rigorous definition through the corresponding state-space
representation is given by \citet{Schmidlkofer2026}.

\begin{table*}
\centering
\caption{Interpretation of the second-order MCARMA damping regimes.
These regimes describe qualitatively different forms of stochastic
variability rather than distinct physical classes of astronomical
sources.}
\label{tab:damping_regimes}
\begin{tabular}{lll}
\hline
Regime & AR structure & Implication for variability \\
\hline
Underdamped ($\zeta<1$)
& Complex-conjugate AR poles
& Damped oscillatory variability with a characteristic frequency\\

Critical ($\zeta=1$)
& Repeated real-valued AR pole
& Transition between oscillatory and non-oscillatory behavior \\

Overdamped ($\zeta>1$)
& Two distinct real-valued AR poles
& Two characteristic decay timescales \\
\hline
\end{tabular}
\end{table*}

We model the stochastic drivers as $k$ correlated Brownian motions,
$\mathbf{B}(t)=(B_1(t),\ldots,B_k(t))^\top$, with
\[
\mathrm{Cov}\{d\mathbf{B}(t)\}=V\,dt,
\qquad
V=D_{\sigma}RD_{\sigma},
\]
where $D_{\sigma}=\mathrm{diag}(\sigma_1,\ldots,\sigma_k)$
contains the band-specific short-term stochastic driving amplitudes
and $R=\{\rho_{jl}\}$ is the correlation matrix of the drivers,
with $\rho_{jl}$ denoting the correlation between bands $j$ and $l$. Admissible
values satisfy $|\rho_{jl}|<1$ for $j\ne l$, with $R$ positive
definite. The stochastic amplitudes $\sigma_j$ control the strength
of the driving fluctuations within individual bands, whereas the
correlations $\rho_{jl}$ introduce dependence between bands.

Following \citet{2020AJ....160..265H} and
\citet{Schmidlkofer2026}, we adopt diagonal AR and MA coefficient
matrices. Each band therefore has its own CARMA$(p,q)$ dynamics,
while dependence across bands arises through the correlated stochastic
drivers. This provides a parsimonious extension of the multivariate
DRW to higher-order band-specific variability; more general
cross-band dynamics can be obtained with non-diagonal coefficient
matrices \citep{tomasson2018}.

For interpreting these band-specific dynamics, we use the roots of
the AR and MA polynomials. The roots of the AR polynomial are referred
to as \emph{AR poles}, whereas the roots of the MA polynomial are
referred to as \emph{MA zeros}. The AR poles determine the characteristic timescales and oscillatory behavior of the process: a real-valued pole corresponds to
exponential relaxation on a characteristic decay timescale, while a
complex-conjugate pair produces damped oscillatory behavior. MA zeros
do not introduce additional persistent dynamical modes; instead, they
modify how stochastic power is distributed across frequency.

For the second-order models considered here, we write the
band-specific AR polynomial as
\[
A_j(z)
=
z^2+2\zeta_j\omega_{n,j}z+\omega_{n,j}^2,
\]
where $\omega_{n,j}$ is the natural angular frequency and
$\zeta_j$ is the damping ratio. The two AR poles are
\begin{align}
\begin{aligned}\nonumber
\lambda_{1,j}
&=
\omega_{n,j}
\left(
-\zeta_j-\sqrt{\zeta_j^2-1}
\right),\\
\lambda_{2,j}
&=
\omega_{n,j}
\left(
-\zeta_j+\sqrt{\zeta_j^2-1}
\right).
\end{aligned}
\end{align}
The damping ratio determines the qualitative form of the dynamics:
$\zeta_j<1$ gives an underdamped process with damped oscillatory
behavior, $\zeta_j=1$ gives a critically damped process with a
repeated real pole, and $\zeta_j>1$ gives an overdamped process with
two distinct real decay rates. We summarize the temporal scale by the
slow characteristic timescale $\tau_j$, defined as the slower
real-pole timescale for $\zeta_j\geq1$ and the damping timescale of
the oscillation envelope for $\zeta_j<1$. For astronomical interpretation, these regimes can be viewed as
qualitatively different forms of stochastic variability, as summarized
in Table~\ref{tab:damping_regimes}; they are not intended to represent
discrete physical classes of AGNs.

For MCARMA$(2,1)$, the band-specific MA polynomial can be written as
\[
M_j(z)=1+b_{1,j}z,
\]
with MA zero $-\omega_{z,j}=-b_{1,j}^{-1}$. We characterize its
location relative to the AR dynamics through the dimensionless
pole--zero ratio
\begin{equation}\label{eq:polezeroratio}
\eta_j=\frac{\omega_{z,j}}{\omega_{n,j}}.
\end{equation}
In the critically damped case, $\eta_j=1$ places the MA zero exactly
at the repeated AR pole, producing pole--zero cancellation, so that the higher-order model becomes mathematically equivalent to a lower-order process. Values near unity
therefore produce near cancellation and weak identifiability. More
generally, the proximity of an MA zero to an AR pole determines the
degree of cancellation. The statistical identifiability of these
features is studied in detail by \citet{Schmidlkofer2026}; here we use
them primarily to interpret the SDSS and LSST results.

The MCARMA$(p,q)$ process has an equivalent linear state-space
representation,
\begin{align}
\begin{aligned}\label{eq:statespace}
d\mathbf{Z}(t)
&=
F\mathbf{Z}(t)\,dt+G\,d\mathbf{B}(t),\\
\mathbf{X}(t)
&=
H\mathbf{Z}(t),
\end{aligned}
\end{align}
where $\mathbf{Z}(t)$ is a $kp$-dimensional latent state vector.
This representation enables likelihood evaluation for irregularly
sampled light curves using Kalman filtering. The construction of the state vector $\mathbf{z}(t)$, the explicit
forms of $F$, $G$, and $H$, and the stationary covariance and
transition distribution are given by \citet{Schmidlkofer2026}.

The frequency-domain behavior of the process is characterized by its
$k\times k$ spectral density matrix $P(\omega)$, where $\omega$
denotes angular frequency. The diagonal entries
$P_{jj}(\omega)$ are the marginal power spectral densities (PSDs),
which describe how variability power is distributed across temporal
frequencies, while the off-diagonal entries $P_{jl}(\omega)$ describe
cross-band spectral dependence. Each marginal PSD has the functional
form of a univariate CARMA$(p,q)$ spectrum, but joint MCARMA fitting
allows information to be shared across correlated bands and can
therefore improve recovery of marginal variability when individual
light curves are sparsely sampled or noisy
\citep{Schmidlkofer2026}.

The multivariate model also characterizes frequency-domain dependence
through the coherence \citep{2014A&ARv..22...72U}, which measures the
strength of the linear relationship between two bands at each
frequency. Following the convention commonly used in the astronomical
literature, we use \emph{coherence} to refer to
$\gamma_{jl}^2(\omega)$, which is often called the magnitude-squared
or squared coherence in the signal-processing and statistical
literature. Under the diagonal coefficient structure adopted here,
\[
\gamma_{jl}^2(\omega)
=
\frac{|P_{jl}(\omega)|^2}
     {P_{jj}(\omega)P_{ll}(\omega)}
=
\rho_{jl}^2.
\]
Thus, the coherence between each pair of bands is constant with
frequency and determined by the correlation between their stochastic
drivers. This frequency independence is a consequence of the
diagonal dynamic structure; more general non-diagonal MCARMA models
can produce frequency-dependent coherence.

The difference between the AR and MA orders determines the
high-frequency behavior of the marginal PSDs:
\[
P_{jj}(\omega)
\propto
|\omega|^{-2(p-q)},
\qquad
|\omega|\rightarrow\infty.
\]
Thus, $p-q=1$ gives the $|\omega|^{-2}$ high-frequency decay of the
DRW, whereas $p-q=2$ gives a steeper $|\omega|^{-4}$ decay,
corresponding to smoother sample paths and stronger suppression of
short-timescale variability. Of the three orders considered here,
MCARMA$(1,0)$ is the multivariate DRW; MCARMA$(2,0)$ introduces
second-order AR dynamics, including the possibility of damped
oscillatory behavior and a steeper high-frequency spectrum; and
MCARMA$(2,1)$ adds an MA component that further modifies the spectral
shape while restoring the $|\omega|^{-2}$ high-frequency decay.

\subsection{Likelihood Evaluation for Astronomical Multiband Data}
\label{sec:likelihood}

The state-space representation in Equation~\eqref{eq:statespace} permits
likelihood evaluation through Kalman filtering while naturally
accommodating irregular observation times, heteroscedastic
measurement errors, and partially observed bands. Let
$t_1<\cdots<t_n$ denote the ordered union of observation times
across all bands, and let $\mathbf{Z}_i=\mathbf{Z}(t_i)$ denote the
latent state at time $t_i$. For $\Delta t_i=t_i-t_{i-1}$, the
continuous-time state equation implies the discrete transition
\[
\mathbf{Z}_i
=
\Phi_i\mathbf{Z}_{i-1}+\boldsymbol{\eta}_i,
\qquad
\Phi_i=\exp(F\Delta t_i),
\]
where
$\boldsymbol{\eta}_i\sim
N_{kp}(\mathbf{0},Q_i)$, with $Q_i$ describing the stochastic
variation accumulated in the latent state over the time interval
$\Delta_i$. The transition covariance $Q_i$ therefore depends on the
time between successive observations, naturally accommodating
irregular sampling. Its explicit form and the stationary
initialization are given in \citet{Schmidlkofer2026}.

Let $\boldsymbol{\mu}=(\mu_1,\ldots,\mu_k)^\top$ denote the
band-specific mean levels, and let
$D_{\boldsymbol{\delta}_i^2}$ be the diagonal matrix containing the
known measurement-error variances at time $t_i$. If all $k$ bands
were observed at every time, the observation equation would be
\[
\mathbf{y}_i
=
\boldsymbol{\mu}
+
H\mathbf{Z}_i
+
\boldsymbol{\epsilon}_i,
\qquad
\boldsymbol{\epsilon}_i
\sim
N_k(\mathbf{0},D_{\boldsymbol{\delta}_i^2}).
\]

Often, only a subset of the bands is observed at a given time. Let $S_i$ denote the $k_i^\ast\times k$ selection matrix that
selects the $k_i^\ast$ bands observed at $t_i$. The observation
equation then becomes
\[
\mathbf{y}_i^\ast
=
S_i\boldsymbol{\mu}
+
S_iH\mathbf{Z}_i
+
\boldsymbol{\epsilon}_i^\ast,
\qquad
\boldsymbol{\epsilon}_i^\ast
\sim
N_{k_i^\ast}
\left(
\mathbf{0},
S_iD_{\boldsymbol{\delta}_i^2}S_i^\top
\right),
\]
where $\mathbf{y}_i^\ast=S_i\mathbf{y}_i$.

At each observation time, the Kalman filter propagates the latent
state according to the elapsed time since the preceding observation
and updates it using only the bands observed at that epoch. The
resulting prediction residuals, known as innovations, and their
covariance matrices provide the Gaussian likelihood for the complete
multiband data set,
\begin{equation*}
\log L(\boldsymbol{\theta};\mathbf{y}^\ast)
=
-\frac{1}{2}
\sum_{i=1}^{n}
\left\{
\log|W_i^\ast|
+
(\mathbf{x}_i^\ast)^\top
(W_i^\ast)^{-1}\mathbf{x}_i^\ast
\right\}
+c,
\end{equation*}
where $\boldsymbol{\theta}$ denotes the collection of model
parameters, $\mathbf{x}_i^\ast$ is the innovation, i.e., the
difference between the observed measurements $\mathbf{y}_i^\ast$
at $t_i$ and their predictions from the preceding data,
$W_i^\ast$ is the covariance matrix of the innovation,
and $c$ is a constant independent of $\boldsymbol{\theta}$. We maximize this likelihood jointly over the
band-specific stochastic parameters and cross-band dependence
parameters. The complete Kalman filtering recursions and likelihood
derivation are given in \citet{Schmidlkofer2026}.

Under the diagonal MCARMA structure considered here, the total
number of free parameters for $k$ bands is
\begin{equation*}
d
=
k(p+q+2)+\frac{k(k-1)}{2}.
\end{equation*}
The first term accounts for the $p$ AR coefficients,
$q$ MA coefficients, one stochastic driving amplitude,
and one mean parameter per band, while the second term
accounts for the pairwise correlations between bands. For the five-band SDSS analysis, this gives
25, 30, and 35 free parameters for MCARMA$(1,0)$,
MCARMA$(2,0)$, and MCARMA$(2,1)$, respectively.

For fixed model orders $p$ and $q$ and number of bands $k$,
the computational cost of the Kalman-filter likelihood
grows linearly with the number of distinct observing epochs.  When only one band is
observed at a given epoch, as is  the case in the LSST
simulations considered below, the Kalman update is univariate. When
several bands are observed nearly simultaneously, as in the SDSS
Stripe~82 data, the corresponding multivariate observation is handled
directly through the selection matrix. The same likelihood formulation
therefore accommodates both observing strategies without interpolating
or synchronizing the light curves.

\subsection{Estimation and Model Selection}
\label{sec:estimation}

We estimate the MCARMA parameters by maximum likelihood using the
parameterization and numerical implementation developed in
\citet{Schmidlkofer2026}. Full details, including parameter
constraints, numerical optimization, uncertainty quantification, and
simulation validation, are provided therein.

Estimation proceeds in two stages. The first stage uses a regularized objective to obtain a numerically stable initial solution when some stochastic features are weakly constrained by the observations. This estimate initializes the second stage, which
maximizes the ordinary, unpenalized likelihood. Thus, the
regularization used during initialization does not enter the final
likelihood, parameter estimates, or model-selection criterion. If no second-stage attempt satisfies the convergence criterion without numerically decreasing the ordinary likelihood, the first-stage estimate is retained, rescored under the ordinary likelihood, and flagged as unconverged. Approximate standard errors are obtained from the local
curvature of the ordinary likelihood when the curvature is positive
definite.

The final ordinary-likelihood fit enables model comparison using the
finite-sample-corrected Akaike information criterion
\citep[AICc,][]{akaike1974new, 10.1093/biomet/76.2.297}.
We select the model with the smallest AICc and use AICc differences
to assess the strength of support relative to alternative model orders.

\section{Numerical Illustrations}
\label{sec:numerical}

We now use the MCARMA framework to investigate stochastic variability
in observed multiband AGN light curves and to assess how well the
resulting behavior can be recovered under realistic LSST observing conditions.
Our numerical analysis proceeds in two parts. We first apply MCARMA to a
carefully selected sample of well-sampled SDSS Stripe~82 AGNs to
characterize the stochastic dynamics and cross-band dependence
supported by existing observations. We then use these empirical
findings to motivate simulations under realistic LSST observing
conditions, examining how cadence and photometric precision affect
the recovery of richer stochastic behavior beyond the multivariate
DRW.

The numerical analyses here are designed around astronomical
interpretation and survey performance rather than general statistical
validation of the MCARMA methodology. Schmidlkofer et al.\
(\citeyear{Schmidlkofer2026}) investigate the statistical properties
of the method through extensive simulation studies, including
model-order recovery, parameter and spectral estimation, weak
identifiability, uncertainty quantification, and computational
performance. Here we focus instead on what the SDSS data reveal about
multiband AGN variability and whether those features remain
recoverable under realistic LSST observing conditions.

\subsection{SDSS Stripe~82 AGNs}
\label{sec:sdss}

We first apply MCARMA to well-sampled multiband AGN light curves from
SDSS Stripe~82. Our primary goals are to determine whether the
multivariate DRW provides an adequate description of the observed
variability, to characterize the additional stochastic structure
supported by higher-order models, and to examine the strength and
wavelength dependence of cross-band variability.

\begin{table*}[t]
\caption{Observing properties and fitted characteristics of the 134
SDSS Stripe~82 AGNs by photometric band. Entries are medians over the sample, with the 25th--75th percentile ranges (\(P_{25}\)--\(P_{75}\)) reported beneath. Measurements
(the number of observations in each band, with their total over the
five bands in the final column), baseline, sampling interval, and
measurement uncertainty describe the observed light curves. Mean
magnitude is each object's time-averaged magnitude in that band,
while the temporal standard deviation measures the observed scatter
about that mean. Comparing the latter with the typical measurement
uncertainty provides a rough indication of how much of the observed
variability may be attributable to measurement error. Stationary S/N
and the slow characteristic timescale are evaluated using the
AICc-selected model. The final column summarizes the five bands.}
\label{tab:sdss}
\setlength{\tabcolsep}{2.5pt}
% generated by fits/process_quasars/sdss_summary_table.py
\centering
\begin{tabular}{lcccccc}
\hline\hline
Band & $u$ & $g$ & $r$ & $i$ & $z$ & all\\
\hline
\multicolumn{7}{l}{\textit{Observing design}}\\
Measurements           & 108 & 110 & 112 & 112 & 111 & 552\\
\quad($P_{25}$--$P_{75}$) & 99--116 & 101--120 & 103--124 & 102--123 & 101--122 & \\
Baseline $T$ (days)       & 3331 & 3331 & 3331 & 3331 & 3331 & 3331\\
\quad($P_{25}$--$P_{75}$)             & 2948--3337 & 2948--3337 & 2949--3337 & 3041--3337 & 2949--3337 & \\
Spacing $\Delta t$ (days) & 2.04 & 2.03 & 2.02 & 2.02 & 2.03 & 2.03\\
\quad($P_{25}$--$P_{75}$)            & 2.01--2.98 & 2.01--2.06 & 2.01--2.05 & 2.01--2.05 & 2.01--2.08 & \\
Mean magnitude         & 20.44 & 20.06 & 19.86 & 19.72 & 19.57 & 19.87\\
\quad($P_{25}$--$P_{75}$)            & 19.77--20.94 & 19.39--20.54 & 19.27--20.32 & 19.10--20.11 & 19.02--20.01 & \\
Median error (mag)     & 0.071 & 0.024 & 0.024 & 0.027 & 0.079 & 0.034\\
\quad($P_{25}$--$P_{75}$)             & 0.044--0.109 & 0.017--0.031 & 0.017--0.032 & 0.019--0.038 & 0.051--0.119 & \\
Temporal sd (mag)      & 0.180 & 0.117 & 0.097 & 0.093 & 0.131 & 0.116\\
\quad($P_{25}$--$P_{75}$)             & 0.122--0.244 & 0.084--0.147 & 0.073--0.126 & 0.069--0.118 & 0.096--0.189 & \\
\hline
\multicolumn{7}{l}{\textit{Fitted at the selected order}}\\
Stationary S/N         & 1.92 & 4.70 & 4.03 & 2.99 & 1.03 & 3.21\\
\quad($P_{25}$--$P_{75}$)             & 1.2--3.1 & 3.2--6.9 & 2.6--5.8 & 2.0--4.4 & 0.7--1.6 & \\
Slow $\tau$ (days)        & 289 & 297 & 337 & 381 & 401 & 337\\
\quad($P_{25}$--$P_{75}$)             & 135--497 & 157--528 & 211--533 & 217--576 & 213--739 & \\
\hline
\end{tabular}
\end{table*}

The observational design nevertheless places important limits on the
temporal scales that can be constrained. Following the
design-based rule of \citet{Schmidlkofer2026}, we treat a
characteristic timescale $\tau$ as informed by the data when
$0.7\,\Delta t\le\tau\le0.2\,T$, where $\Delta t$ is the median
sampling interval and $T$ is the baseline. The upper limit requires the baseline to cover about five
$e$-folding times (the time over which the correlation decays
to $1/e$ of its initial value), and the lower limit excludes
timescales much shorter than the typical spacing between
measurements.  With the sample medians $\Delta t=2.03$ days and
$T=3331$ days (Table~\ref{tab:sdss}), the practically informative
range of characteristic timescales is approximately 1.4--666 days.
These limits are heuristic consequences of the observing design, not
formal identifiability bounds. We therefore interpret fitted dynamical quantities relative
to these observational limits below. Unless otherwise stated,
timescales and frequencies reported for the Stripe~82 sample
are in the observed frame; the PSDs in
Figure~\ref{fig:sdss.psd} instead use rest-frame frequency.

\subsubsection{Data and sample selection}
\label{sec:sdss.data}

We analyze the SDSS Stripe~82 five-band $ugriz$ light curves of 9,258
spectroscopically confirmed quasars from the SDSS DR7 quasar catalog
compiled by \citet{macleod2012}, using their original irregular
sampling and heteroscedastic measurement uncertainties. We first
require at least 400 individual band measurements summed over
the five filters, reducing the sample to 386 candidate objects. After linear detrending, we apply a per-band stationarity screening procedure to
exclude light curves exhibiting behavior inconsistent with the
stationary stochastic models considered here. This removes 252
objects and leaves a final sample of 134 AGNs.

An epoch is one Stripe~82 imaging run, in which the five
filters are observed within minutes of each other; it corresponds to
one row of the \citet{macleod2012} light-curve files. The retained
objects contain between 82 and 143 observing epochs, with a median of 115, corresponding to between 400 and 688 individual
band measurements and a median of 552. The median observational
baseline is 3331 days. Although the five filters are not observed
strictly simultaneously, measurements within an epoch are typically
separated by only a few minutes; the median separation is
approximately 4.8 minutes, and 89\% of the epochs contain measurements
in all five bands. This near-coincident multiband sampling provides
particularly strong information about cross-band dependence.

Table~\ref{tab:sdss} summarizes the observing properties and fitted
characteristics of the Stripe~82 sample. Photometric quality varies
appreciably across bands. The median measurement uncertainty is smallest in \(g\) and \(r\) and largest in \(u\) and \(z\).
The mean magnitudes of the selected sample range from approximately
19.5 in $z$ to 20.4 in $u$, with the $gri$ bands concentrated near
magnitude 20. These brightnesses provide a useful empirical reference
for the LSST simulations below, where magnitude 19.5 represents a
brightness scale comparable to the selected Stripe~82 sample and
magnitude 22.0 probes a substantially fainter regime.

\subsubsection{Selected model orders}
\label{sec:sdss.select}

AICc selects MCARMA$(1,0)$ for 25 of the 134 AGNs, MCARMA$(2,0)$ for
4, and MCARMA$(2,1)$ for 105. Thus, a second-order model is selected
for 109 objects (81.3\%), and MCARMA$(2,1)$ alone is selected for 105
(78.4\%). Within the MCARMA family considered here, these results
provide strong empirical evidence that models more flexible than the
multivariate DRW are needed for detailed characterization of the
variability in most of these Stripe~82 AGNs.

Incomplete numerical convergence (i.e., failure to satisfy our final optimization convergence criteria) does not weaken this central
conclusion. MCARMA$(1,0)$ converges for every object, whereas an
unconverged higher-order fit provides a lower bound on its attainable
maximized log-likelihood and therefore an upper bound on its AICc. All
109 objects currently selecting a higher-order model already favor it
over the converged MCARMA$(1,0)$ fit, so further optimization cannot
reverse any of these classifications in favor of the multivariate
DRW. Among the 25 objects currently selecting MCARMA$(1,0)$, 15 have
an unconverged higher-order competitor that could in principle improve
enough to change the classification. The observed 81.3\% higher-order
fraction is therefore a conservative lower bound on the prevalence of
stochastic structure beyond the multivariate DRW within the candidate
family considered here. Incomplete numerical convergence can still affect the specific distinction between MCARMA\((2,0)\) and MCARMA\((2,1)\).

The strength of the model preference differs substantially between
objects selecting the simpler and higher-order models. Among the 105
objects selecting MCARMA$(2,1)$, the median AICc difference relative
to the next-best fitted model is 40.3, and only 11 objects have an AICc
difference smaller than 10. In contrast, among the 25 objects selecting
MCARMA$(1,0)$, the median difference is 4.7 and 22 objects have differences
smaller than 10. Thus, when a higher-order model is selected, its
advantage over the competing fitted models is generally much stronger
than the corresponding advantage when the multivariate DRW is
selected.

These model-selection results should not be interpreted as implying
that the underlying physical processes are literally
MCARMA$(2,1)$. Rather, within the candidate family considered here,
the observed light curves generally support stochastic structure
beyond that represented by the multivariate DRW. Moreover, evidence
for a higher-order stochastic model does not imply that every
parameter or characteristic timescale of that model is individually
well constrained. Such weakly constrained higher-order structure also
motivates our two-stage estimation procedure. The regularized first
stage stabilizes optimization and provides reliable starting values,
whereas the subsequent unpenalized refinement maximizes the ordinary
likelihood on which AICc is based. The refinement changes the selected
order relative to the regularized first-stage solution for 36 of the
134 AGNs, demonstrating the importance of separating numerical
stabilization from final model selection. These considerations motivate emphasizing the inferred PSDs,
cross-band coherence, and robust distinctions between the
multivariate DRW and higher-order structure, rather than
assigning physical meaning to every fitted higher-order parameter.

\subsubsection{Characteristic timescales and spectral structure}
\label{sec:sdss.timescales}

The fitted AR poles provide a compact description of the
characteristic timescales and oscillatory behavior in each band. Real-valued poles correspond to exponential
decay components, whereas complex-conjugate poles correspond to damped
oscillatory modes. For a second-order model with two real-valued poles, we
refer to the associated longer and shorter decay scales as the slow and
fast timescales. In MCARMA$(2,1)$, the additional MA zero reshapes the
spectrum rather than introducing another characteristic
decay timescale. An MA zero
that approaches an AR pole can nearly cancel that pole's contribution,
creating the near-cancellation regime described in
Section~\ref{sec2.1}. These distinctions are important when interpreting
the fitted timescales below.

The selected second-order models are predominantly overdamped,
consistent with the single-band CARMA$(2,1)$ fits of
\citet{2022ApJ...936..132Y,2025ApJ...992..130Y}. Because the AR
and MA coefficient matrices are diagonal, each band has its own pair
of AR poles and, for MCARMA$(2,1)$, its own MA zero. These roots are
separate parameters for each band; the bands are linked only through
the correlations of their stochastic drivers. Among
the 545 band fits belonging to the 109 objects selecting a second-order
model, 522 have two real-valued AR roots, and 93 of the 109
objects have real-valued roots in all five bands. Thus, the higher-order variability supported by the Stripe~82
data is predominantly characterized by multiple characteristic
decay timescales rather than damped oscillatory behavior.

The slower decay timescale generally increases from the bluer to the
redder bands. Its median values are approximately 289, 297, 337, 381,
and 401 days in $u$, $g$, $r$, $i$, and $z$, respectively, with a
pooled median of 337 days. These values lie within the practically
informative range for the typical Stripe~82 light curve. The upper
tail, however, is substantially less constrained: the 90th percentile
is approximately 1016 days, and 153 of the 670 band fits place the
slow timescale beyond the corresponding object's practical upper
boundary ($0.2\,T$ for that object, or 666 days at the median
baseline). Such long fitted timescales should therefore be regarded as extrapolations beyond the well-constrained temporal range rather than
precise measurements.

The second-order models also exhibit substantially shorter
characteristic decay timescales. The median fitted fast decay
timescale is approximately 0.034 days, and 81.1\% of these
fitted timescales fall below the practical lower boundary of
approximately 1.4 days implied by the observational design.
For MCARMA$(2,1)$, the MA zero of band $j$ lies at
$-1/m_j$, where $m_j$ is the $j$th diagonal entry of $M_1$,
so $m_j$ is the MA-zero timescale. In 55 of the 525 band fits, $m_j$ is below
$10^{-3}$ days (and exactly zero in 35), so the MA term effectively
drops out and that band follows the CARMA$(2,0)$ form. Across the
remaining 470, the median MA-zero timescale is approximately 0.91
days, with the 25th and 75th percentiles at 0.38 and 4.2 days, respectively. These short-timescale quantities should therefore not generally be interpreted as well-resolved physical timescales.
Instead, the additional second-order and MA terms provide
the flexibility needed to represent departures from the simple DRW
spectral shape over the frequencies to which the data are sensitive.

The strong preference for MCARMA$(2,1)$ does not appear
to be driven by near pole--zero cancellation, a weak-identifiability
regime investigated by \citet{Schmidlkofer2026}. Only 1 of the 525 MCARMA$(2,1)$ band fits places the MA
zero $z_j$ within 10\% of the nearest AR pole $\lambda_j$, that is,
satisfies $|\lambda_j-z_j|<0.1\,|\lambda_j|$. Thus, the prevalence of MCARMA$(2,1)$ in the Stripe~82 sample cannot
generally be attributed to this particular weak-identifiability
regime.

\begin{figure*}[tp]
\centering
\includegraphics[width=\textwidth]{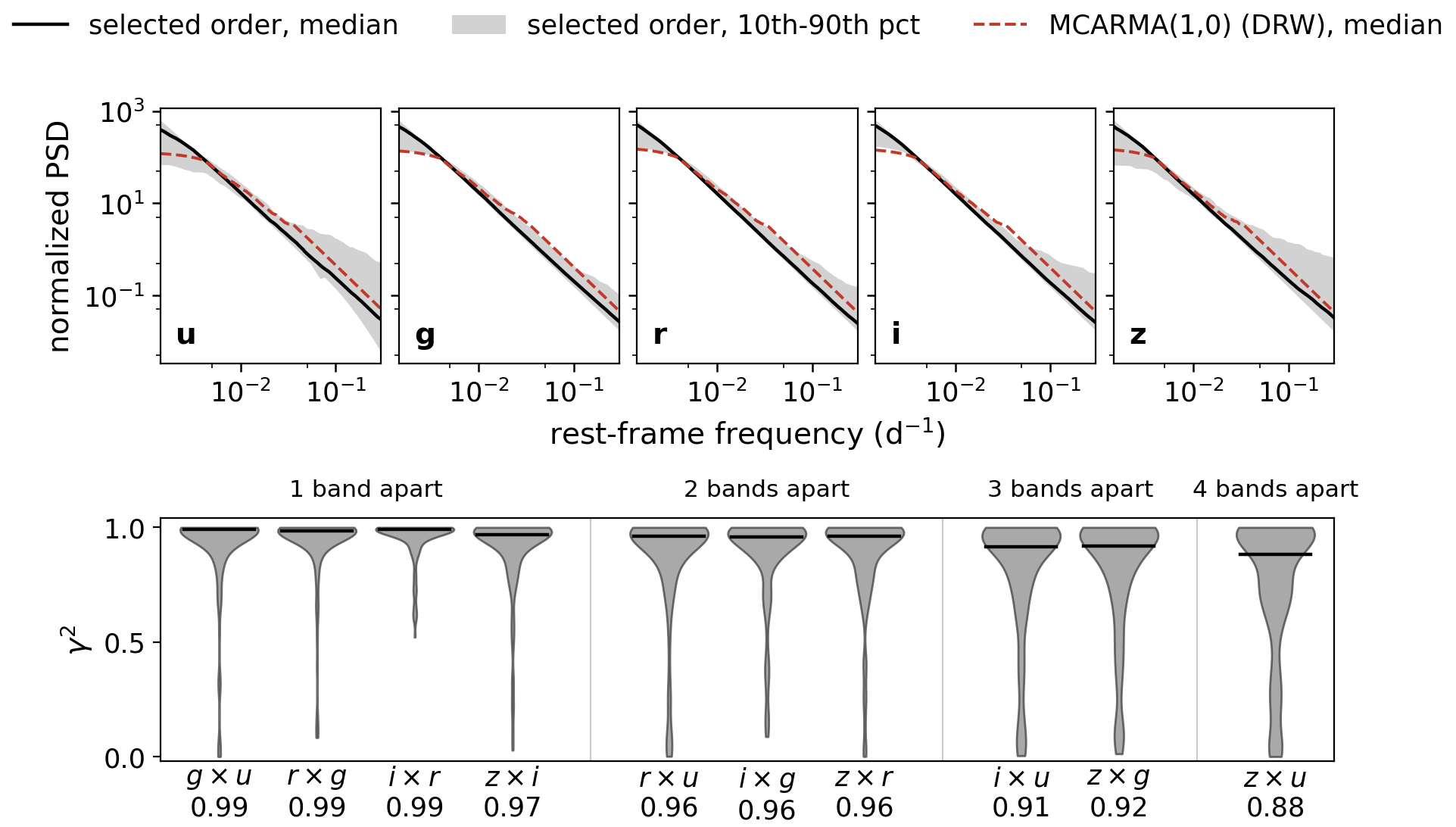}
\caption{Population-level spectral and cross-band dependence structure
of the 134 SDSS Stripe~82 AGNs. Top panels show the marginal power
spectral densities for the five photometric bands as a function of
rest-frame frequency, over the range that lies inside the observable
window of every object. Each object's PSD
is first normalized to unit area over the plotted frequency range so
that the panels compare spectral shape rather than amplitude. The solid
curve is the pointwise median of the 134 normalized PSDs at each
frequency, and the shaded region is the corresponding pointwise tenth
to ninetieth percentile range. Thus, the solid curve is a population
summary rather than the spectrum of a selected representative object.
The dashed curve is the pointwise median of the same objects'
MCARMA$(1,0)$ (DRW) fits, normalized in the same way. The bottom panel
shows violin distributions of the estimated coherence
($\gamma^2$) across the 134 objects for each of the ten band pairs,
grouped by wavelength separation; violin width represents the relative
density of objects, the horizontal bar marks the median, and the median
is also printed under each pair label. Cross-band coherence is strongest
between neighboring filters and decreases with increasing wavelength
separation. Because some estimated correlations lie near the boundary
of the admissible correlation space ($|\rho_{jl}|$ close to
1), the coherence distributions are
intended to show the overall pattern of cross-band dependence rather
than precise estimates of individual coherence values.}
\label{fig:sdss.psd}
\end{figure*}

\subsubsection{Cross-band dependence}
\label{sec:sdss.coherence}

Figure~\ref{fig:sdss.psd} summarizes the fitted marginal PSDs and
cross-band dependence for the 134 AGNs. The top panels show the
population distribution of marginal spectral shapes in each
photometric band as a function of rest-frame frequency
$f_{\rm rest}=f_{\rm obs}(1+z)$, using the redshifts of
\citet{macleod2012}. Because each object's PSD is normalized to unit area over the plotted frequency range, these panels compare \emph{shape}
rather than variability amplitude: relatively greater power toward the
left corresponds to variability concentrated on longer timescales,
whereas bends and changes in slope indicate characteristic stochastic
frequencies and departures from a single DRW-like spectral shape. The
black curve is the pointwise median of the 134 normalized PSDs at each
frequency, and the gray envelope is the corresponding pointwise central
80\% range. Thus, the black curve summarizes the population spectral
shape rather than representing any individual object, while the width
of the envelope reflects object-to-object diversity. The
dashed curve is the pointwise median of the same objects'
multivariate DRW fits, normalized in the same way.
Relative to it, the selected-order median carries more power at low
frequency, by a factor of about 3 at the lowest plotted frequency, and
less above about $5\times10^{-3}$~d$^{-1}$, by up to a factor of about
2 near $0.04$~d$^{-1}$. Over the highest plotted decade both medians
fall roughly as $f^{-2}$. The selected higher-order models therefore
place relatively more variability on the longest timescales than the
DRW fits do. The lowest plotted frequencies correspond to roughly one
cycle per baseline for the highest-redshift objects, so the size of
the low-frequency excess is less certain than its presence.

The bottom panel summarizes the fitted coherence for each of
the ten band pairs using violin plots, with the pairs grouped by
wavelength separation. At a given coherence value, the width of a violin is proportional to the density of objects with fitted values
near that level; a narrow tail therefore represents relatively few
objects, whereas a broad region indicates a concentration of the
sample. The number printed under each pair label is the median coherence. The strong concentration of the violins near unity shows that most objects
have very strong cross-band coupling, while the progressively broader
downward tails for more widely separated filters reveal increasing
object-to-object variation and weaker dependence.

The fitted dependence exhibits a clear wavelength pattern. Cross-band
dependence is strongest between neighboring filters and decreases as
the separation between filters increases. The median absolute
stochastic-driver correlations for filter separations of one through
four are approximately 0.993, 0.980, 0.959, and 0.940, corresponding
to coherences of approximately 0.99, 0.96, 0.92, and 0.88. Thus,
variability is strongly shared across all five bands, while the fitted
dependence weakens systematically with increasing wavelength
separation.

These fitted correlations should not be interpreted as
precise estimates of the underlying cross-band dependence.
Approximately 27\% of adjacent-band correlations exceed 0.999,
and 88 of the 134 objects have at least one fitted correlation
close to the boundary of the admissible correlation space,
that is, with $|\rho_{jl}|>0.999$. The robust empirical conclusion is therefore the pattern of very strong adjacent-band dependence that decreases with
wavelength separation, rather than the precise numerical values of
the individual correlations.

As a sensitivity check, we recomputed the summary statistics after
restricting the sample first to the 85 objects whose selected order
cannot be overturned by an unconverged competing order and then to the
62 objects for which all three candidate orders converged. The median
PSDs and the decline of coherence with wavelength separation remain
qualitatively unchanged under both restrictions; the 85-object PSD
summaries in particular track those from the full sample closely.
Because excluding objects based on numerical convergence may
introduce selection bias, we retain all 134 objects for the
primary sample summaries in Figure~\ref{fig:sdss.psd}
and use the restricted analyses only as robustness checks.

\subsubsection{Interpretation and motivation for the LSST study}
\label{sec:sdss.reading}

The Stripe~82 analysis illustrates both the scientific information
provided by the MCARMA framework and the limitations imposed by the
observational design. The data provide strong evidence in favor of
models more flexible than the multivariate DRW for most objects and
show a clear pattern of strong cross-band dependence that decreases
with wavelength separation. At the same time, several fitted
dynamical quantities lie near or beyond the temporal range that can be
reliably constrained by the Stripe~82 cadence and baseline. In
particular, the fast AR timescale is unresolved for most second-order
fits, while the longest slow timescales extend beyond the effective
baseline constraint.

The model-selection result should therefore be distinguished from
precise identification of every fitted dynamical parameter. The
preference for higher-order models indicates that the observed
variability contains stochastic structure not adequately represented
by the multivariate DRW within the candidate family considered here;
it does not imply that every fitted AR pole or MA zero corresponds to a
well-resolved physical timescale. This distinction is consistent with
the simulation study of \citet{Schmidlkofer2026}, which shows
that successful model-order selection need not imply equally precise
recovery of every individual dynamical parameter.

The Stripe~82 results provide both the motivation and an observational
reference point for the LSST study below. Higher-order stochastic
structure is strongly supported in most of the selected SDSS AGNs, but
these conclusions are obtained from a particularly favorable sample:
the objects are relatively bright, the five bands are observed nearly
simultaneously, and each light curve contains many multiband epochs.
LSST will probe AGN variability under substantially different
conditions. Its observations are irregular and staggered across
filters, and its greater depth will extend variability studies to
sources considerably fainter than those dominating the selected
Stripe~82 sample. We therefore use realistic LSST observing sequences
to ask whether the richer multiband stochastic behavior identified
above remains recoverable as temporal sampling and photometric
precision change.

\subsection{Rubin LSST Cadence Study}
\label{sec:lsst}

We investigate how the recovery of MCARMA stochastic structure changes
under realistic observing conditions expected for Rubin LSST. We
compare the relatively sparse WFD cadence with the much denser DDF
cadence and consider two source brightness levels, allowing the effects
of temporal sampling and photometric precision to be examined
separately. The study asks which features of the underlying stochastic
dynamics remain recoverable, how model-order and spectral recovery
depend on cadence and source brightness, and whether joint multivariate
modeling continues to improve inference when individual bands are
irregularly and sparsely sampled.

\subsubsection{Simulation design}
\label{sec:num.sweep.design}

We consider ten generative settings, summarized in
Table~\ref{tab:spec.sweep}. Three  settings use
MCARMA$(1,0)$ with $u$-band characteristic timescales of 30, 100,
and 300 days. Six additional settings use second-order dynamics:
MCARMA$(2,0)$ and MCARMA$(2,1)$ in the underdamped,
critically damped, and overdamped regimes.   The final setting is a critically damped
MCARMA$(2,1)$ process designed to produce near pole--zero
cancellation, a weakly identifiable configuration discussed in
Section~\ref{sec2.1} and \citet{Schmidlkofer2026}.

\begin{table}[t]
\caption{Generative settings for the Rubin LSST cadence study.
Within each setting, the characteristic timescale increases with
wavelength from $\tau_u$ to $\tau_y=2\tau_u$, with the intermediate
bands spaced linearly between these endpoints. For the second-order
models, $\zeta$ denotes the damping ratio, with $\zeta<1$, $\zeta=1$,
and $\zeta>1$ corresponding to underdamped, critically damped, and
overdamped dynamics, respectively. For MCARMA$(2,1)$, $\eta$ denotes
the pole--zero ratio defined in Equation~\eqref{eq:polezeroratio}. Both quantities,
when applicable, are shared across bands. The final row represents
the weakly identifiable near pole--zero cancellation setting.}
\label{tab:spec.sweep}
\centering
\begin{tabular}{rrrrr}
\hline\hline
Model & Damping &  & & \\
order & regime & $\tau_u$ (d) & $\zeta$ & $\eta$\\
\hline
$(1,0)$ & ---      & 30  & ---  & ---\\
$(1,0)$ & ---      & 100 & ---  & ---\\
$(1,0)$ & ---      & 300 & ---  & ---\\
$(2,0)$ & underdamped       & 150 & 0.15 & ---\\
$(2,0)$ & critically damped & 150 & 1.00 & ---\\
$(2,0)$ & overdamped        & 150 & 3.90 & ---\\
$(2,1)$ & underdamped       & 150 & 0.15 & 2.5\\
$(2,1)$ & critically damped & 150 & 1.00 & 2.5\\
$(2,1)$ & overdamped        & 150 & 3.90 & 2.5\\
\hline
$(2,1)$ & critically damped & 150 & 1.00 & 1.2\\
\hline
\end{tabular}
\end{table}

\begin{figure*}[t]
\centering
\includegraphics[width=\textwidth]{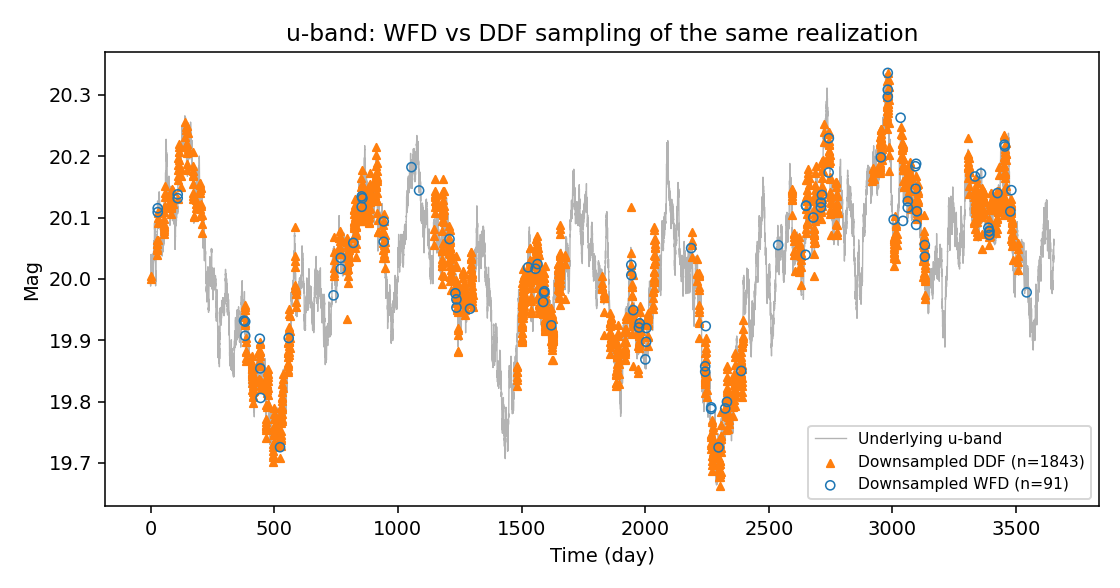}
\caption{Illustration of the Rubin LSST observing sequences used in the
simulation study. The same underlying $u$-band MCARMA$(1,0)$
realization (gray curve) is sampled using representative
Wide-Fast-Deep (WFD; circles) and Deep Drilling Field (DDF; triangles)
observing sequences, containing 91 and 1843 observations,
respectively. The comparison illustrates the substantially denser
temporal sampling of the DDF observations while holding the underlying
stochastic variability fixed.}
\label{fig:sampling}
\end{figure*}

For the six primary second-order settings, the $u$-band slow
characteristic timescale $\tau_u$ is 150 days. Within each setting,
the characteristic timescale increases with wavelength from
$\tau_u$ to $\tau_y=2\tau_u$, with the intermediate $g$, $r$, $i$,
and $z$ bands spaced linearly between these endpoints. We use
$\zeta=0.15$, 1, and 3.9 to represent underdamped, critically damped,
and overdamped dynamics, respectively. These choices span qualitatively different second-order behavior:
for $\tau_u=150$ days, the overdamped setting has a fast
$u$-band AR-pole timescale of 2.6 days (5.1 days in $y$),
the critically damped setting has a repeated AR pole
corresponding to a timescale of 150 days, and the underdamped
setting has an oscillation period of approximately 143 days.

For the primary MCARMA$(2,1)$ settings, we use a common pole--zero
ratio $\eta=2.5$. The corresponding $u$-band MA-zero timescales are
7.8, 60, and 9.0 days in the overdamped, critically damped, and
underdamped settings, respectively. The additional weakly identifiable
setting uses $\zeta=1$ and $\eta=1.2$, for which the MA-zero timescale
is 125 days and hence lies close to the 150-day repeated AR-pole
timescale. This configuration produces the near pole--zero
cancellation described in Section~\ref{sec2.1}. The stochastic
drivers have correlation $\rho_{jl}=0.9^{|j-l|}$ between bands $j$
and $l$, ordered as $u,g,r,i,z,y$, so neighboring bands are strongly
correlated and the dependence decreases smoothly with wavelength
separation.

The simulations begin from densely sampled (30-second cadence)
continuous-time MCARMA realizations and then apply realistic Rubin
observing sequences and brightness-dependent photometric
uncertainties. This construction separates the intrinsic stochastic
process from the information loss introduced by cadence and
measurement noise.

Each simulated process is sampled using realistic Rubin LSST observing
sequences at 100 locations uniformly spaced in the WFD footprint and
the six DDFs. For each generative setting, the same underlying
continuous-time MCARMA process is sampled under the WFD and DDF
designs, helping isolate the effect of cadence from stochastic
realization-to-realization variability. Figure~\ref{fig:sampling}
illustrates this construction directly: the dense gray curve shows the
underlying continuous-time realization, while the WFD and DDF points
show how differently the same stochastic trajectory is observed. In
the representative $u$-band example, the WFD and DDF sequences contain
91 and 1843 observations, respectively. The exact number and timing of
observations vary with field and band, and the state-space likelihood
handles these irregular and partially observed multiband sequences
directly without interpolation or synchronization.

Each observing sequence is considered at apparent magnitudes 19.5 and
22.0, which we refer to as the bright and faint settings. The intrinsic
stochastic variability amplitude is held fixed, while the Rubin
photometric uncertainty changes with source brightness and observing
conditions. Consequently, the two brightness levels differ in
\emph{variability S/N}, the ratio of the stationary intrinsic
variability amplitude to the per-epoch photometric uncertainty. This
is distinct from the usual flux-based detection S/N. The bright
setting is comparable to the characteristic brightness of the
selected Stripe~82 sample, whereas the faint setting represents a
substantially fainter LSST regime.

Measurement errors are generated according to the adopted Rubin
photometric-error model at the corresponding magnitude
\citep{ivezi2019lsst} and are treated as known heteroscedastic
uncertainties when fitting the simulated light curves. Each simulated
data set is fitted with all three candidate orders,
MCARMA$(1,0)$, MCARMA$(2,0)$, and MCARMA$(2,1)$, using the estimation
procedure described in Section~\ref{sec:estimation}, and model order
is selected using AICc computed from the ordinary log-likelihood at
the retained fit. Thus, the design permits direct comparisons between
stochastic regimes, sparse WFD and dense DDF sampling, and bright and
faint photometric precision.

Because only six DDF sequences are available compared with 100 WFD
locations, aggregate DDF recovery fractions are based on substantially
fewer independent cadence realizations. Differences involving
individual DDF settings should therefore be interpreted with
appropriate caution.

\subsubsection{Model-order recovery}
\label{sec:num.sweep.order}

Table~\ref{tab:recovery.sweep} summarizes correct-order recovery
across the WFD and DDF simulations for the bright and faint source
settings. Recovery depends strongly on the form of the underlying
stochastic variability as well as on cadence and photometric
precision. Across many of the simulated regimes, higher-order
dynamics are reliably distinguished from the multivariate DRW, but
recovery deteriorates substantially in particular second-order
settings, especially for faint sources under sparse WFD sampling.
Thus, whether LSST can recognize stochastic structure beyond the
DRW depends not only on how well a source is observed, but also on
how distinctly that structure is expressed in its light curve.

Reduced photometric precision generally makes it more difficult
to distinguish higher-order structure from the multivariate DRW,
although the effect varies substantially among the generative
settings. Under WFD sampling, the effect is particularly
pronounced for the overdamped MCARMA$(2,0)$ and the underdamped
and critically damped MCARMA$(2,1)$ settings. Dense DDF sampling
generally improves recovery at both brightness levels, although
the improvement is not uniform across settings. Individual DDF
recovery rates should nevertheless be interpreted cautiously
because each setting contains only six DDF observing sequences.

\begin{table}[t]
\centering
\caption{Correct-order selection rates across the Rubin LSST simulation
study under WFD and DDF sampling. Model order denotes the generative
MCARMA$(p,q)$ order. For the second-order models, ``under,''
``critical,'' and ``over'' denote the underdamped, critically damped,
and overdamped regimes, respectively. The near pole--zero cancellation setting is reported separately and
excluded from the mean. The mean is an unweighted summary of the nine
primary simulation settings and should not be interpreted as a
population-averaged LSST recovery rate.}
\label{tab:recovery.sweep}
% GENERATED by fits/21sim/recovery_tables.py -- do not edit
% arm: _TSUNIDLe_jaxse (2026-09-30), complete 2026-10-06 with the two split DDF
% refits ((1,0) tau 300 faint, (2,1) critical bright).
\begin{tabular}{rrrcccc}
\hline\hline
Model & Damping & $\tau_u$ & \multicolumn{2}{c}{WFD} & \multicolumn{2}{c}{DDF}\\
\cline{4-5}\cline{6-7}
order & regime & (d) & bright & faint & bright & faint\\
\hline
(1,0) & --- & 30 & 0.92 & 0.95 & 0.83 & 1.00\\
(1,0) & --- & 100 & 0.72 & 0.85 & 1.00 & 0.67\\
(1,0) & --- & 300 & 0.81 & 0.92 & 1.00 & 0.50\\
(2,0) & under & 150 & 0.94 & 0.96 & 1.00 & 1.00\\
(2,0) & critical & 150 & 0.90 & 0.91 & 1.00 & 1.00\\
(2,0) & over & 150 & 0.87 & 0.14 & 1.00 & 0.67\\
(2,1) & under & 150 & 1.00 & 0.60 & 1.00 & 1.00\\
(2,1) & critical & 150 & 0.72 & 0.29 & 0.67 & 0.33\\
(2,1) & over & 150 & 0.99 & 0.79 & 0.83 & 0.83\\
\hline
\multicolumn{3}{l}{\emph{unweighted mean}} & 0.87 & 0.71 & 0.93 & 0.78\\
\hline
\multicolumn{7}{l}{\emph{Near pole--zero cancellation, excluded from the mean}}\\
(2,1) & critical & 150 & 0.33 & 0.42 & 0.00 & 0.00\\
\hline
\end{tabular}
\end{table}

The critically damped MCARMA$(2,1)$ setting has the lowest recovery
among the primary second-order settings. Because $\zeta=1$ corresponds to critical damping, marking the
transition between complex and real-valued AR poles, we checked
whether this difficulty is specific to the exact choice
$\zeta=1$ by repeating the experiment at $\zeta=0.90$, 0.95,
1.00, 1.05, and 1.10, holding $\tau_u=150$ days and
$\eta=2.5$ fixed (Appendix~\ref{app:lsst.zeta}). Recovery does not exhibit an isolated
minimum at $\zeta=1$: in all four cadence and brightness combinations
it is highest at $\zeta=0.90$ and lowest at $\zeta=1.05$ or 1.10.
For bright WFD sources, recovery decreases from 0.93 at $\zeta=0.90$
to 0.55--0.60 near $\zeta=1$ and 0.24--0.31 above it, with almost
every mis-selected source classified as MCARMA$(1,0)$. Over the same
range, the median log-likelihood gain of MCARMA$(2,1)$ over
MCARMA$(1,0)$ decreases from 24.0 to about 10, becoming comparable
to or smaller than the approximately 12-nat gain needed to offset
the AICc penalty for the 12 additional parameters.

This behavior indicates that the difficulty is not confined to the
single critically damped configuration. As $\zeta$ approaches unity
from below, the complex-conjugate AR poles coalesce, weakening the
distinctive oscillatory structure of the second-order process. Above unity, the two AR poles become distinct and real-valued,
corresponding to two characteristic decay timescales, while
the faster pole moves progressively closer to the MA zero
in relative terms, increasing partial pole--zero cancellation. In both cases,
the observable distinction from a lower-order DRW can become weak.
Thus, selection of a DRW model does not necessarily imply
intrinsically first-order stochastic dynamics; genuinely higher-order
variability can become observationally difficult to distinguish from
the DRW.

The MCARMA$(1,0)$ settings illustrate a complementary distinction
between selecting the correct model order and accurately estimating
its characteristic timescale. Under bright WFD sampling, correct-order
recovery remains high even for the 300-day setting, although
estimation of the corresponding long-timescale dynamics becomes less
accurate. This is consistent with the simulation results of
\citet{Schmidlkofer2026}: successful order selection does not by itself
imply that all timescales of the selected model are well constrained
by the observational design.

\subsubsection{Marginal spectral and cross-band recovery}
\label{sec:num.sweep.accuracy}

Model-order recovery alone does not describe how accurately the fitted
stochastic dynamics are reconstructed. We therefore use the marginal
PSD as a scientifically interpretable summary of the joint parameter
estimation and coherence as a complementary summary of cross-band
dependence. We use the shape-normalized spectral error (SNSE) of
\citet{Schmidlkofer2026} to select representative marginal PSD
recoveries. SNSE compares recovered and generative PSD shapes after
normalizing each spectrum by its own value at the lowest evaluated
frequency. SNSE is zero for exact recovery of the spectral shape, and
smaller values indicate better recovery. Detailed parameter- and
spectral-accuracy results, including the formal SNSE definition and
full error distributions, are given in
Appendix~\ref{app:lsst.accuracy}.

Figure~\ref{fig:psd.lsst}(a) shows representative marginal PSD
recoveries selected by their SNSE. Each panel gives the generative
$g$-band marginal PSD for one simulated source together with the
multivariate and univariate recoveries, with the three generative
orders stacked vertically and the two cadences and two source
magnitudes forming a square within each order. All three curves are
divided by their own value at $f_0$, which is the normalization SNSE
applies before integrating. The source drawn in a panel is the one at
the median SNSE of its cell, so the panels show typical recovery rather
than the best or worst case; the twelve representative sources and the
cell-median SNSEs of the multivariate and univariate fits are reported
in Table~\ref{tab:lsst.psd.objects}.

The difference in PSD recovery accuracy between the multivariate
and univariate fits is most pronounced under WFD sampling. Under
DDF sampling, both recoveries closely follow the generative spectrum
in all three orders, and the univariate fit performs at least as well
as the multivariate fit in these representative cases. Under WFD
sampling, however, the univariate fit misplaces power in every order,
and in the same direction each time: too much power above the break in
the faint MCARMA$(1,0)$ panel, an overshoot at the resonance in the
faint MCARMA$(2,0)$ panel, and, in the faint MCARMA$(2,1)$ panel, a
resonant peak at roughly one-third of the generative frequency. The
multivariate fit places the peak correctly in the latter case, although
it falls below the generative spectrum at the highest frequencies.
With roughly 600--750 observations spread over six bands, individual
WFD bands contain substantially less information than the joint
multiband data. The cell medians in
Table~\ref{tab:lsst.psd.objects} show that the advantage of joint
fitting is particularly pronounced for the higher-order models, for
which the univariate SNSE is approximately two to four times larger.
By contrast, the dense DDF sampling allows individual bands to recover
their marginal spectra accurately without borrowing information across
bands; joint modeling remains essential, however, for inference on
cross-band dependence.

\begin{figure*}[!t]
\centering
\includegraphics[width=\textwidth]{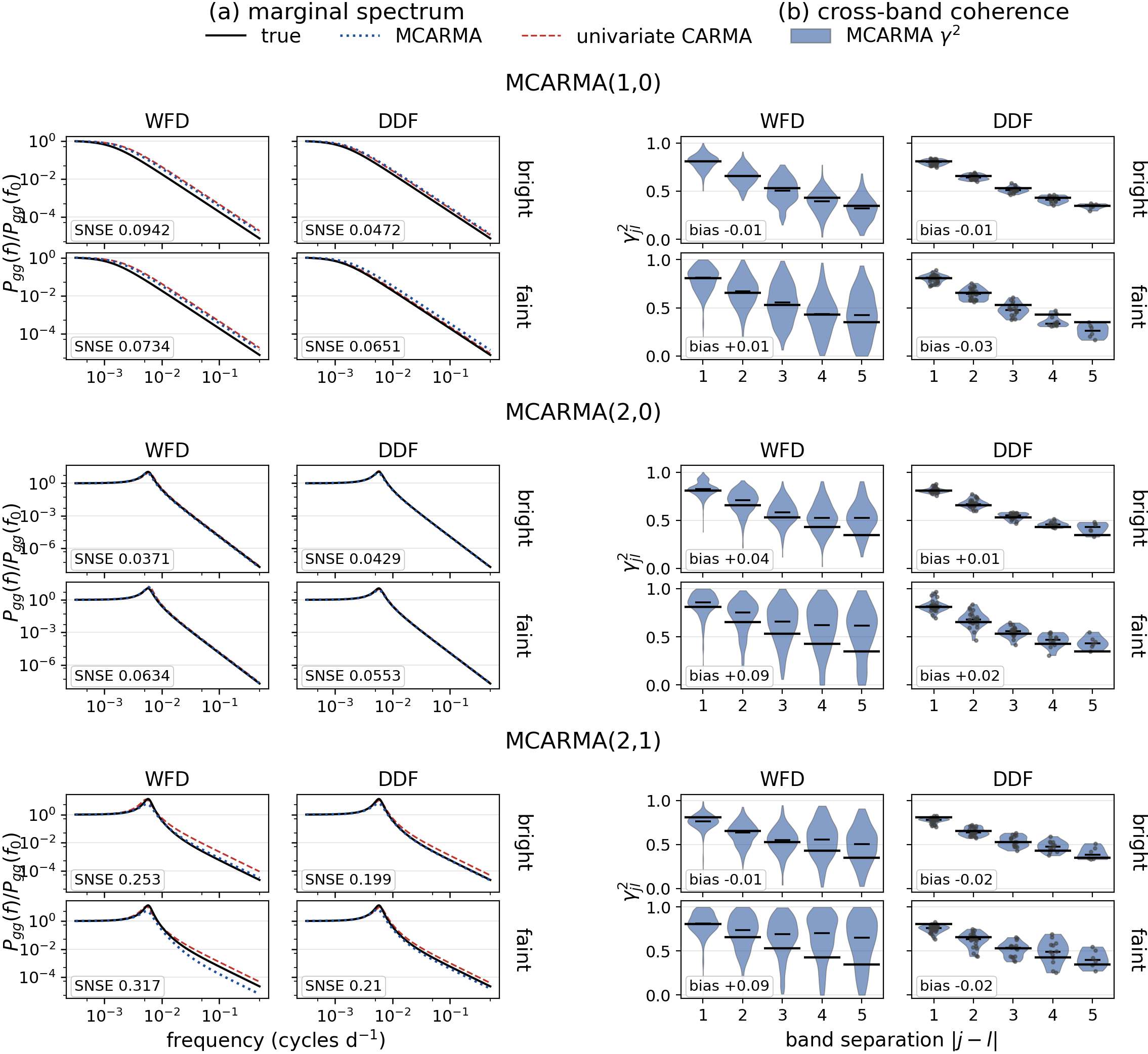}
\caption{Recovery of marginal power spectra (a) and cross-band
coherence (b) for representative simulated Rubin LSST sources. Both
halves share the same layout: the three blocks correspond to the
generative MCARMA orders; within each block, columns show the WFD and
DDF cadences and rows show the bright and faint source magnitudes.
(a)~Recovered and generative $g$-band marginal power spectra. The
black solid curves show the generative PSDs, while the blue dotted and
red dashed curves show the PSDs recovered from joint MCARMA and
separate univariate CARMA fits, respectively. Each curve is normalized
by its value at the lowest evaluated frequency $f_0$, so the
comparison emphasizes spectral shape rather than overall amplitude.
For each combination of MCARMA order, cadence, and source magnitude,
we show the source with the median shape-normalized spectral error
(SNSE) from the joint MCARMA fit, with its SNSE reported in the
corresponding panel. The twelve representative sources and the
cell-median SNSEs for the multivariate and univariate fits are
reported in Appendix~\ref{app:lsst.accuracy}. (b)~Fitted coherence
$\widehat\gamma^2_{jl}$ against band separation $|j-l|$, shown for
every source in the cell rather than for the single source drawn in
(a). Under the diagonal coefficient structure adopted here the
coherence does not depend on frequency, so each fit contributes one
value per band pair, and a violin pools the $6-|j-l|$ pairs at a given
separation over all sources in the cell: 100 sky locations under WFD
and 6 deep drilling fields under DDF, whose individual band pairs are
also plotted. The wide black bar spanning each violin marks the
generative coherence $0.9^{2|j-l|}$, the shorter black line inside the
body is the median of the fitted values, and the number reported in
each panel is the median signed difference from the generative value,
pooled over separations. The univariate CARMA comparator of (a)
estimates each band separately and therefore has no counterpart in
(b).}
\label{fig:psd.lsst}
\end{figure*}

\begin{table}[t]
\centering
\caption{Representative sources shown in
Figure~\ref{fig:psd.lsst}. $N$ is the total number of observations
across all six bands. SNSE is the $g$-band shape-normalized spectral
error of the multivariate fit for the representative source. The last
two columns give the corresponding cell-median SNSEs for the
multivariate and per-band univariate fits, providing a reference for
how representative the displayed source is and for the relative
spectral-recovery accuracy of the two approaches.}
\label{tab:lsst.psd.objects}
{\footnotesize\setlength{\tabcolsep}{4pt}
% GENERATED by fits/21sim/lsst_psd_grid.py -- do not edit
% arm: _TSUNIDLe_jaxse (two-stage PMLE-MLE, 5 per cent stage-1 diagonal loading, univariate companions fitted)
\begin{tabular}{llrrrr}
\hline\hline
Order & Sampling & $N$ & SNSE & \multicolumn{2}{c}{Cell median SNSE} \\
\cline{5-6}
 & & & & MCARMA & Univariate \\
\hline
(1,0) & WFD bright & 717 & 0.0942 & 0.0922 & 0.1160 \\
(1,0) & DDF bright & 17813 & 0.0472 & 0.0422 & 0.0301 \\
(1,0) & WFD faint & 665 & 0.0734 & 0.0731 & 0.1370 \\
(1,0) & DDF faint & 14332 & 0.0651 & 0.0560 & 0.0182 \\
\hline
(2,0) & WFD bright & 733 & 0.0371 & 0.0366 & 0.0815 \\
(2,0) & DDF bright & 14522 & 0.0429 & 0.0410 & 0.0130 \\
(2,0) & WFD faint & 709 & 0.0634 & 0.0633 & 0.1760 \\
(2,0) & DDF faint & 21489 & 0.0553 & 0.0508 & 0.0195 \\
\hline
(2,1) & WFD bright & 624 & 0.2530 & 0.2510 & 0.5860 \\
(2,1) & DDF bright & 14522 & 0.1990 & 0.1910 & 0.0963 \\
(2,1) & WFD faint & 612 & 0.3170 & 0.3150 & 1.2200 \\
(2,1) & DDF faint & 17813 & 0.2100 & 0.2040 & 0.0898 \\
\hline
\end{tabular}}
\end{table}

Cross-band dependence is summarized in
Figure~\ref{fig:psd.lsst}(b). Under the diagonal coefficient-matrix structure, the coherence
between a pair of bands is independent of frequency,
so each fit yields a single coherence value per pair,
with generative coherence $0.9^{2|j-l|}$. Each violin includes all sources in the
corresponding simulation setting and pools band pairs having the same
separation $|j-l|$. Thus, unlike the spectral panels, which display
one representative source, the coherence panels summarize the full
ensemble of simulated sources.

The recovered coherence generally follows the expected decline with
increasing wavelength separation, but its accuracy depends strongly on
cadence. Under DDF sampling, the distributions are narrow and the
median error remains below 0.10 at every separation. Under WFD
sampling, however, the distributions are substantially broader and the
second-order models tend to overestimate coherence at larger
separations. For example, at $|j-l|=5$, the faint MCARMA$(2,0)$ and
MCARMA$(2,1)$ settings both have median recovered coherence of 0.62,
compared with the generative value of 0.35. By contrast, adjacent
bands, with generative coherence 0.81, are recovered to within 0.06
of the generative value in every setting.

The WFD bias is therefore concentrated among bands furthest apart in
wavelength, where the true dependence is weakest: sparse sampling can
lead the fitted model to infer more shared variability than is
supported by the generative process. The univariate fits in
Figure~\ref{fig:psd.lsst}(a) have no counterpart here because
independent single-band analyses cannot estimate cross-band
dependence.

Detailed accuracy results in Appendix~\ref{app:lsst.accuracy} support
these representative comparisons. Across the nine primary settings, the geometric mean ratio of
univariate to multivariate SNSE is 2.56 and 76.8 for the bright
and faint WFD samples, respectively, and 1.78 and 5.97 for the
corresponding DDF samples; the two ratios for faint sources
reflect outright failure of the single-band fits in $u$,
the noisiest band, rather than a uniform advantage. In DDF this failure
occurs in a single setting, without which the faint ratio is 2.16. Estimation of the characteristic frequencies associated with
the AR poles shows the same general dependence on cadence
and photometric precision, becoming more difficult for faint
sources under sparse WFD sampling.

\subsubsection{Near pole--zero cancellation}
\label{sec:num.nonid}

The final MCARMA$(2,1)$ setting examines whether denser sampling or
improved photometric precision can resolve the weak identifiability
associated with near pole--zero cancellation. Here $\zeta=1$ and
$\eta=1.2$, so the MA zero lies close to the repeated AR pole, as
described in Section~\ref{sec2.1}.

Model-order recovery remains poor under both observing strategies. As
shown in Table~\ref{tab:recovery.sweep}, correct-order recovery is
0.33 for bright sources and 0.42 for faint sources under WFD sampling,
and 0.00 for both brightness levels under DDF sampling. Under WFD the
spectral deterioration is also pronounced: for bright sources the
median SNSE is 51.1, compared with a maximum of 3.6 across the other
generative settings. Under DDF the bright-source median SNSE is 1.5,
below the maximum of 2.0 across the other settings, while
for faint sources it is 4.5 against a maximum of 0.34 elsewhere. For
bright DDF sources the generative-order fit therefore reproduces the
spectrum, yet AICc still does not select it.

The persistence of this difficulty under substantially denser DDF
sampling is consistent with the near-cancellation mechanism studied
in detail by \citet{Schmidlkofer2026}. As an MA zero approaches an AR
pole, their contributions nearly cancel, flattening the corresponding
spectral feature and making the resulting PSD resemble that of a
lower-order process. Thus, improved cadence and photometric precision
can mitigate observation-limited failures, but they need not resolve
structural weak identifiability. In this regime, selecting
MCARMA$(1,0)$ does not necessarily imply genuinely first-order
stochastic dynamics. We therefore report the near-cancellation
setting separately from the aggregate performance summaries.

\subsubsection{Summary of the LSST study}
\label{sec:lsst.summary}
The LSST simulations show that recovery of multiband stochastic
structure depends jointly on the underlying stochastic dynamics,
photometric precision, and observing cadence. Higher-order dynamics
are reliably distinguished from the multivariate DRW across many of
the simulated settings, but recovery deteriorates when their
characteristic variability features become less distinct from
lower-order behavior, particularly for faint sources under sparse WFD
sampling. The damping-ratio sensitivity analysis shows that this
difficulty is not confined to a single finely tuned configuration:
genuinely higher-order variability can appear DRW-like over a range
of underlying dynamics. Moreover, correct model-order selection does
not by itself imply that all characteristic stochastic features are
well constrained. Denser sampling can mitigate observation-limited
failures, but does not necessarily resolve structural weak
identifiability such as near pole--zero cancellation.

The simulations also show that the benefit of multivariate modeling
persists under realistic survey conditions. Joint fitting improves
marginal spectral recovery across the primary settings under both WFD
and DDF sampling. Together with the strong cross-band dependence
found in Stripe~82, this result demonstrates that the information
shared across filters can be scientifically useful not only for
measuring cross-band dependence itself, but also for recovering the
stochastic dynamics within individual bands.

\section{Conclusion}
\label{sec:conclusion}

We have applied the MCARMA framework developed by
\citet{Schmidlkofer2026} to investigate stochastic variability beyond
the multivariate DRW in multiband AGN
observations. Our analysis addresses two complementary astronomical
questions: whether richer stochastic behavior is supported by
well-sampled SDSS observations and whether such behavior can remain
recoverable under the substantially different cadence and photometric
conditions expected from Rubin LSST.

The SDSS Stripe~82 analysis provides strong empirical motivation for
moving beyond the multivariate DRW. Among the 134 well-sampled AGNs
retained after screening, 109 (81.3\%) favor higher-order MCARMA
models, with substantial rather than marginal support for the
preferred higher-order model in most objects. This fraction is conservative with respect to possible numerical nonconvergence. The selected second-order models are
predominantly overdamped, indicating that the additional variability
structure is generally characterized by multiple characteristic
decay timescales rather than damped oscillatory behavior. The fitted
marginal PSDs exhibit diverse stochastic behavior across objects
and bands, while cross-band dependence is generally strong and
decreases with increasing wavelength separation. These patterns
remain stable under more restrictive convergence-based sensitivity
analyses. Thus, the additional structure supported by these data
is not simply a refinement of a single characteristic timescale:
it involves both richer within-band stochastic dynamics and
systematic dependence across wavelength.

The realistic LSST simulations clarify when such structure can be
recovered observationally. Recovery depends not only on photometric
precision and observing cadence, but also strongly on the form of the
underlying stochastic variability. Higher-order dynamics are reliably
distinguished from the multivariate DRW across many of the simulated
settings, but recovery deteriorates when their characteristic
variability features become less distinct from lower-order behavior,
particularly for faint sources under sparse WFD sampling. The
damping-ratio sensitivity analysis shows that this difficulty can
persist over a range of underlying dynamics rather than being
confined to a single finely tuned configuration. Moreover, correct
model-order selection does not by itself imply that all
characteristic stochastic features are well constrained. Denser
sampling can mitigate observation-limited failures, but does not
necessarily resolve structural weak identifiability such as near
pole--zero cancellation. Thus, selection of a DRW model does not
necessarily imply intrinsically first-order stochastic dynamics. At
the same time, joint multivariate fitting improves marginal spectral
recovery under realistic survey conditions, demonstrating the value
of information shared across bands when individual light curves are
information-limited.

The prevalence of higher-order behavior in the SDSS sample raises
questions that extend beyond determining which stochastic model best
describes an individual AGN. A natural next step is to investigate
how the inferred characteristic timescales and PSD shapes vary with
wavelength and with physical source properties such as black-hole
mass, luminosity, accretion state, and redshift. Relationships of this kind have been extensively investigated
using DRW parameters; MCARMA makes it possible to examine whether
additional characteristic frequencies, spectral slopes, or
moving-average features carry complementary physical information. Analyses of larger AGN samples may also reveal whether
the diverse PSD shapes vary systematically with AGN
properties or correspond to distinct regimes of
stochastic variability.

The multiband nature of the model provides a second direction for
physical interpretation. The strong dependence inferred between
neighboring SDSS bands, together with its general decline with
wavelength separation, suggests that the dependence structure itself
should be treated as an observable rather than merely as a device for
improving marginal estimation. Examining how cross-band dependence
changes with wavelength, temporal frequency, and AGN properties could
help distinguish variability that is shared broadly across the
emitting region from behavior that becomes increasingly
band-specific. The present structured MCARMA model provides a
deliberately simple starting point, with frequency-independent
coherence. More general cross-spectral models will be particularly
useful if future observations reveal frequency-dependent
cross-band coherence or other systematic multiband structure
not captured by this assumption.

Our WFD--DDF comparison also has implications for how future
variability studies use LSST observations. Dense DDF sampling can
substantially improve the recovery of stochastic structure when
inference is limited by sparse cadence or reduced photometric
precision, although the magnitude of this improvement depends
strongly on the underlying variability regime. Dense fields may
therefore provide especially valuable samples for detailed
characterization of PSD shape, characteristic timescales, and
cross-band dependence, while the much larger WFD sample can provide
complementary population information for stochastic features that
remain distinguishable under sparse sampling. Combining these regimes
offers an opportunity to use dense samples to establish and calibrate
detailed variability relationships that can subsequently be
investigated across the much larger WFD population.

Realizing this opportunity at LSST scale will require moving beyond
object-by-object likelihood optimization. Simulation-based inference
\citep{2020PNAS..11730055C,papamakarios2021normalizing} provides one
promising route by amortizing inference over simulated MCARMA
observations with realistic cadences and measurement uncertainties.
More broadly, hierarchical or population-level approaches could
directly characterize distributions of stochastic timescales,
spectral features, and cross-band dependence and relate them to AGN
properties while accounting for heterogeneous observing conditions.
Such analyses would shift the role of stochastic variability models
from descriptions of individual sources toward statistical probes of
the AGN population.

Overall, the main implication of this work is not simply that a
higher-order stochastic model fits many of the well-sampled SDSS AGNs
better than the multivariate DRW. Rather, the combination of diverse
marginal spectral structure, systematic cross-band dependence, and
the regime-dependent recoverability of these features under realistic
LSST conditions opens the possibility of using multiband stochastic
variability itself as a population-level observable. Rubin LSST will
provide the sample size, wavelength coverage, and range of observing
densities needed to investigate how this variability structure
changes across the AGN population and what those changes reveal about
the physical processes driving AGN variability.

\section*{Acknowledgements}
We thank Lishan Shi for helpful discussions on the application
of MCARMA models to observational data.
H. Tak gratefully acknowledges partial support from the
Brain Pool Fellowship Program 2026--27 of the National Research
Foundation of Korea (RS-2026-25551883).

\software{\texttt{mcarma} v0.1.2 \citep{mcarma2026},
\url{https://github.com/ischmidls/mcarma}}

\appendix

\section{Additional Results for the Rubin LSST Cadence Study}
\label{app:lsst.accuracy}

This section provides additional parameter- and spectral-accuracy
results underlying the Rubin LSST study in
Section~\ref{sec:lsst}. The main text emphasizes model-order recovery
and the scientifically interpretable consequences of joint parameter
estimation through marginal PSDs and cross-band coherence. Here we
report the full accuracy summaries across the simulation ensemble,
followed by a check of how the critically damped MCARMA$(2,1)$
setting depends on the damping ratio.

\subsection{Detailed parameter and spectral recovery}
\label{app:lsst.recovery}

We use the shape-normalized spectral error (SNSE) of
\citet{Schmidlkofer2026} to compare recovered and generative marginal
PSD shapes. SNSE compares the two spectra on the linear scale after
each has been divided by its own value at the lowest evaluated
frequency, so that only spectral shape enters. For band $j$, let
$P_{jj}(f)$ and $\widehat P_{jj}(f)$ denote the generative and
estimated marginal PSDs. Writing $f_0$ for the lowest evaluated
frequency and $\bar P_{jj}(f)=P_{jj}(f)/P_{jj}(f_0)$, and likewise
$\widehat{\bar P}_{jj}$ for the estimate,
\begin{equation}
\operatorname{SNSE}_j
=
\frac{\displaystyle\int_{f_0}^{f_1}
\left[\widehat{\bar P}_{jj}(f)-\bar P_{jj}(f)\right]^2\,df}
{\displaystyle\int_{f_0}^{f_1}
\left[\bar P_{jj}(f)\right]^2\,df}.
\label{eq:snse.app}
\end{equation}
SNSE is evaluated on a logarithmically spaced grid running from
$f_0=10^{-3.5}$ to $f_1=10^{-0.3}$ cycles per day and is averaged
over all six $ugrizy$ bands for the aggregate comparisons below.
The same grid is used for every data set, so SNSE values are comparable
across generative settings. Smaller values indicate better recovery.

\begin{figure*}[tp]
\centering
\includegraphics[width=\textwidth]{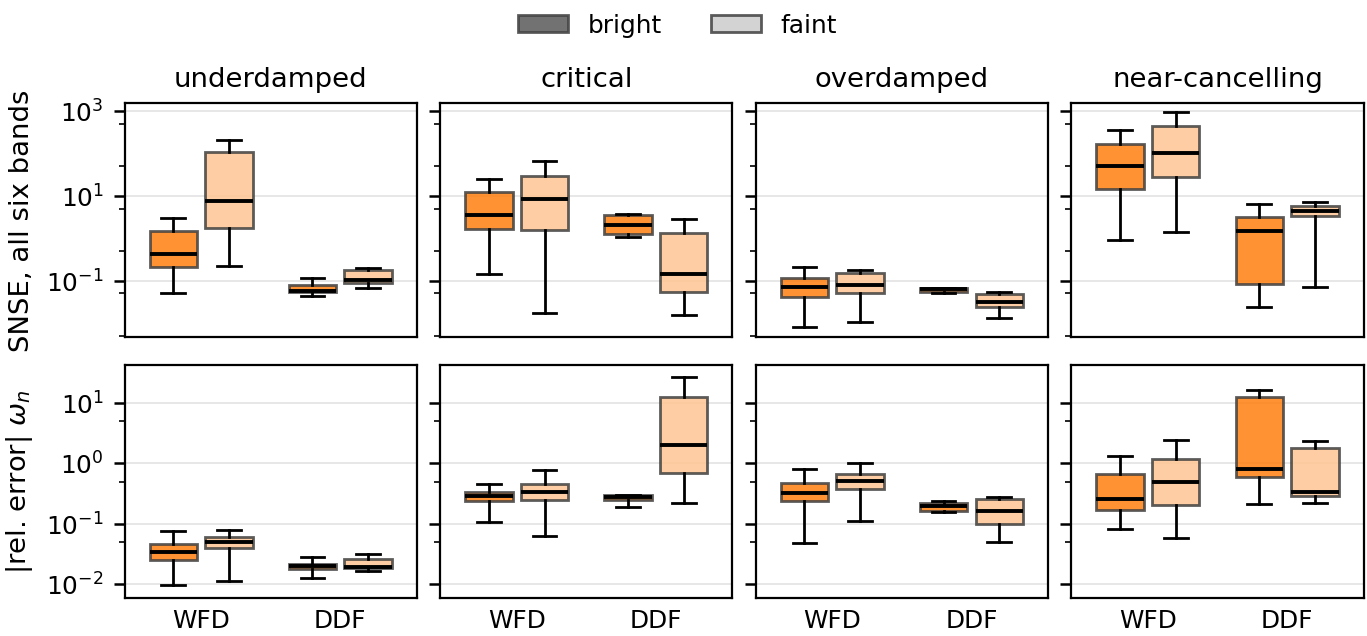}
\caption{Parameter and spectral accuracy for the three regular
MCARMA$(2,1)$ settings and the near pole--zero cancellation setting.
The top row shows shape-normalized spectral error (SNSE) and the bottom
row shows relative error in the estimated characteristic frequency
$\omega_n$, both evaluated at the generative order. Within each panel,
the left pair of boxes is WFD and the right pair DDF sampling. Paired
boxes compare source magnitudes 19.5 and 22.0, the lighter box of each
pair being the faint source; hue identifies the generative order, here
MCARMA$(2,1)$ throughout. The WFD results use 100 footprint locations
per setting, whereas the DDF results use the six Deep Drilling Fields.
Smaller values indicate more accurate recovery.}
\label{fig:sweep.focus}
\end{figure*}

Figure~\ref{fig:sweep.focus} focuses on the three regular
MCARMA$(2,1)$ settings and the near pole--zero cancellation setting,
comparing spectral and characteristic-frequency recovery under WFD
and DDF sampling. The characteristic frequency of band $j$ is
the natural frequency $\omega_{n,j}=|\lambda_{1,j}\lambda_{2,j}|^{1/2}$,
the geometric mean of the magnitudes of its two AR poles, which is
defined whether the poles are real or complex; for MCARMA$(1,0)$ it is
$1/\tau_j$. Its relative error $|\hat\omega_{n,j}-\omega_{n,j}|/\omega_{n,j}$
is summarized by the median over the six bands. Under WFD sampling, estimation generally becomes
less accurate for fainter sources, although the magnitude of this
effect depends on the underlying dynamics. The denser DDF observations
generally improve both characteristic-frequency and spectral recovery,
especially when sparse sampling limits inference. The near-cancellation
setting remains substantially more difficult, illustrating that denser
sampling can mitigate observation-limited failures but does not
necessarily resolve structural weak identifiability.

\begin{figure*}[tp]
\centering
\includegraphics[width=\textwidth]{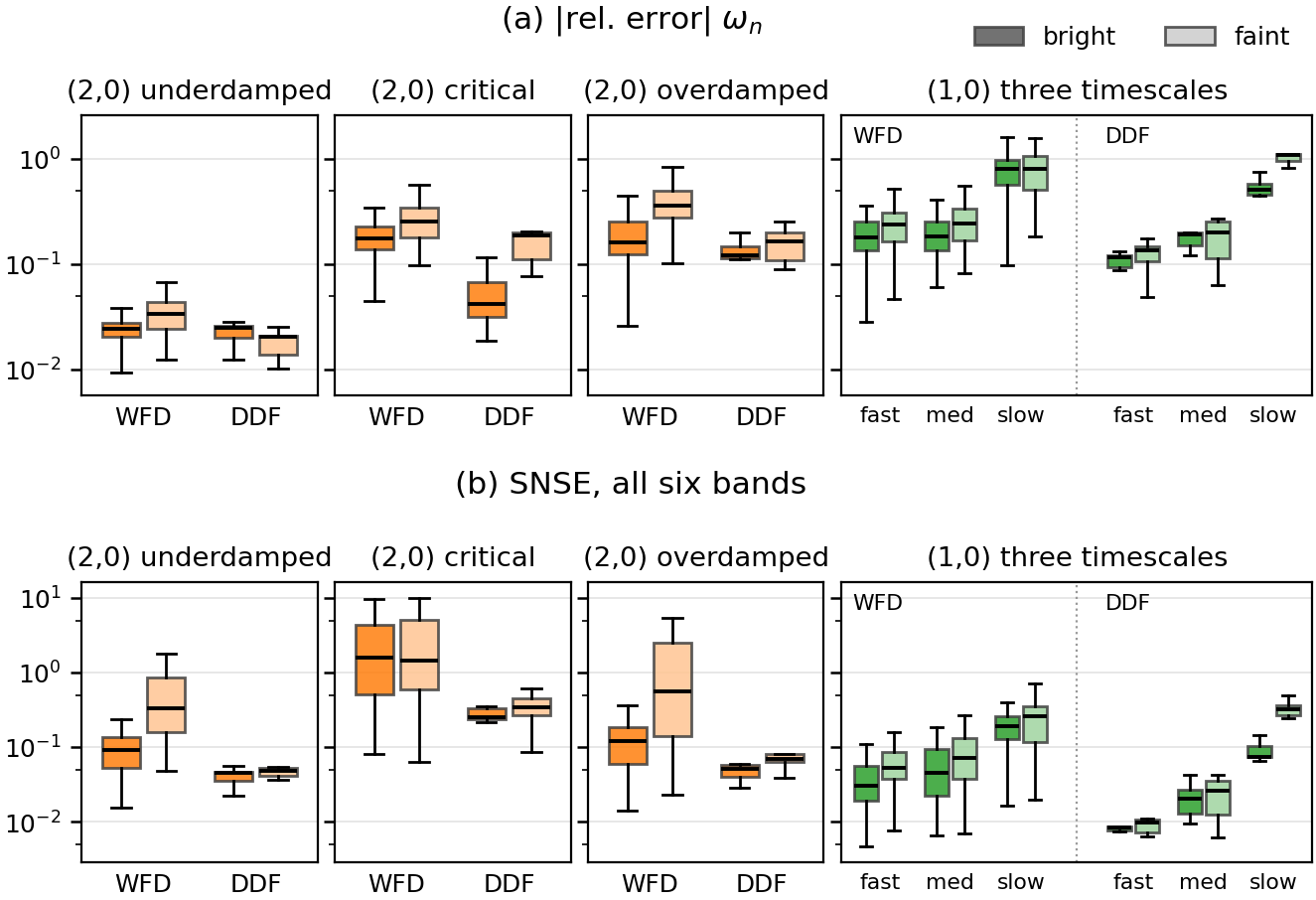}
\caption{Characteristic-frequency accuracy (a) and marginal
spectral accuracy (b) for the additional Rubin LSST simulation settings
at the generative order. Panel (a) shows relative error in the fitted
characteristic frequency and panel (b) shows shape-normalized spectral
error (SNSE; \citealt{Schmidlkofer2026}), with WFD sampling on the
left and DDF sampling on the right of each panel, and bright and
faint source settings compared within each pair. The first three columns correspond
to the underdamped, critically damped, and overdamped MCARMA$(2,0)$
settings, respectively; the fourth column contains the three
MCARMA$(1,0)$ characteristic-timescale settings. Like
Figure~\ref{fig:sweep.focus}, hue identifies the generative order, the
lighter box of each pair is the faint source, and within each of (a)
and (b) all panels share one vertical scale. Smaller values indicate
more accurate recovery.}
\label{fig:sweep.relerr}
\end{figure*}

The same dependence on sampling, photometric precision, and intrinsic
stochastic dynamics appears in the remaining MCARMA$(2,0)$ and
MCARMA$(1,0)$ settings. Figure~\ref{fig:sweep.relerr}(a) shows that
characteristic-frequency recovery depends jointly on temporal sampling,
photometric precision, and the location of the characteristic
frequency relative to the range informed by the observing design.
Dense DDF sampling generally improves recovery relative to WFD,
although the magnitude of the improvement varies across stochastic
regimes. The MCARMA$(1,0)$ results also show that correct model-order
recovery does not necessarily imply equally accurate estimation of a
long characteristic timescale.

Figure~\ref{fig:sweep.relerr}(b) gives the corresponding marginal
spectral-accuracy results. The critically damped MCARMA$(2,0)$ setting
has substantially larger SNSE than the other two damping regimes under
both cadences, while the slowest MCARMA$(1,0)$ setting has larger
spectral error than the two shorter-timescale settings. These
differences are not apparent from model-order recovery rates alone and
further illustrate why model selection and recovery of the detailed
stochastic spectrum should be assessed separately.

Joint multivariate modeling is particularly beneficial when individual
bands are information-limited. Across the ten WFD settings, the
multivariate fit has smaller characteristic-frequency error in eight
settings for both the brighter and the fainter sources.
Across the nine primary settings, the geometric mean ratio of
univariate to multivariate SNSE is 2.56 and 76.8 for the bright and
faint WFD samples, respectively, and 1.78 and 5.97 for the
corresponding DDF samples. The two faint figures reflect outright
failure of the single-band fits in $u$, the noisiest band, rather than
a uniform advantage across bands; in DDF the failure is confined to
the underdamped MCARMA$(2,1)$ setting, without which the faint DDF
ratio is 2.16. These aggregate comparisons complement the
representative marginal PSDs in Figure~\ref{fig:psd.lsst} and the
cell-median results in Table~\ref{tab:lsst.psd.objects} of the main
text.

\subsection{Sensitivity of the critically damped MCARMA(2,1) setting to the damping ratio}
\label{app:lsst.zeta}

\begin{table}[!t]
\centering
\caption{Sensitivity of correct-order recovery for the
MCARMA$(2,1)$ setting ($\tau_u=150$ days, $\eta=2.5$) to the damping
ratio $\zeta$, under WFD (100 sequences) and DDF (6 sequences)
sampling. The last two columns give the median log-likelihood gain
$\Delta\ell=\ell_{(2,1)}-\ell_{(1,0)}$ for WFD; a gain of about 12 nats is needed
to offset the AICc penalty for the 12 additional parameters of MCARMA$(2,1)$.}
\label{tab:zeta.sens}
% arm: _TSUNIDLe_zsens (2026-10-04), same estimator as _TSUNIDLe; tags 21_z090_sep..21_z110_sep
\begin{tabular}{rcccccc}
\hline\hline
 & \multicolumn{2}{c}{WFD} & \multicolumn{2}{c}{DDF} & \multicolumn{2}{c}{WFD median $\Delta\ell$}\\
\cline{2-3}\cline{4-5}\cline{6-7}
$\zeta$ & bright & faint & bright & faint & bright & faint\\
\hline
0.90 & 0.93 & 0.65 & 0.83 & 0.83 & 24.0 & 18.6\\
0.95 & 0.55 & 0.13 & 0.67 & 0.83 & 14.5 & 9.5\\
1.00 & 0.60 & 0.11 & 0.67 & 0.33 & 14.6 & 10.8\\
1.05 & 0.24 & 0.06 & 0.33 & 0.17 & 9.9 & 10.2\\
1.10 & 0.31 & 0.10 & 0.17 & 0.50 & 11.1 & 9.4\\
\hline
\end{tabular}
\end{table}

Table~\ref{tab:zeta.sens} repeats the MCARMA$(2,1)$ configuration
used for the critically damped setting of Table~\ref{tab:spec.sweep}
at five values of $\zeta$, with $\tau_u=150$ days and $\eta=2.5$
unchanged. New densely sampled
realizations were generated for each value and downsampled with the
visit times and per-visit $5\sigma$ depths of the WFD and DDF
sequences used for that setting in the main study, so only the
generative process differs. Each cell holds the first 100 WFD
sequences and all six DDF sequences, and the fits use the same
estimator as the main study.

In every column recovery is highest at $\zeta=0.90$ and lowest at
$\zeta=1.05$ or 1.10. Below 1 the two AR poles are complex with
modulus equal to the natural frequency, so the MA zero stays a factor
$\eta$ away from them. As $\zeta$ approaches unity from below,
however, the imaginary parts of the complex poles shrink and the pair
approaches a repeated real pole, weakening the distinctive oscillatory
behavior of the second-order process even though the pole--zero ratio
remains fixed. Above 1 the poles split into two distinct real poles,
and the faster one moves toward the MA zero in relative terms,
increasing partial pole--zero cancellation. Exact cancellation would
occur at $\zeta=(\eta+\eta^{-1})/2=1.45$, so the settings above 1 move
toward the near-cancellation configuration of
Section~\ref{sec:num.nonid}. Almost all misselected bright sources
select MCARMA$(1,0)$, while misselections are split between
MCARMA$(1,0)$ and MCARMA$(2,0)$ for faint WFD sources. The
$\zeta=1$ row is an independent realization of the setting in
Table~\ref{tab:recovery.sweep}, where WFD recovery is 0.72 and 0.29,
so differences of that size between neighboring rows are within
realization variability.

\bibliography{reference}
\bibliographystyle{aasjournalv7}

\end{document}